# Astrobiologically Interesting Stars within 10 parsecs of the Sun

Gustavo Porto de Mello
Eduardo Fernandez del Peloso[1]
Luan Ghezzi

Observatório do Valongo, Universidade Federal do Rio de Janeiro
Ladeira do Pedro Antônio, 43, Rio de Janeiro, 20080-090 Brazil

[1]Present address: Observatório Nacional, Rua Gen. José Cristino 77,
Rio de Janeiro, 20921-400 Brazil.

**ABSTRACT**
The existence of life based on carbon chemistry and water oceans relies upon planetary properties, chiefly climate stability, and stellar properties, such as mass, age, metallicity and Galactic orbits. The latter can be well constrained with present knowledge. We present a detailed, up-to-date compilation of the atmospheric parameters, chemical composition, multiplicity and degree of chromospheric activity for the astrobiologically interesting solar-type stars within 10 parsecs of the Sun. We determine their state of evolution, masses, ages and space velocities, and produce an optimized list of candidates that merit serious scientific consideration by the future space-based interferometry probes aimed at directly detecting Earth-sized extrasolar planets and seeking spectroscopic infrared biomarkers as evidence of photosynthetic life. The initially selected stars number 33 solar-type within the population of 182 stars (excluding late M-dwarfs) closer than 10 pc. A comprehensive and detailed data compilation for these objects is still essentially lacking: a considerable amount of recent data has so far gone unexplored in this context. We present 13 objects as the nearest "biostars", after eliminating multiple stars, young, chromospherically active, hard X-ray emitting stars, and low metallicity objects. Three of these "biostars", HD 1581, 109358 and 115617, closely reproduce most of the solar properties and are considered as premier targets. We show that approximately 7% of the nearby stars are optimally interesting targets for exobiology.

## I - INTRODUCTION

The goal of this paper is to present a list of stars, among the stellar population within 10 parsecs of the Sun, which best qualify as adequate hosts to habitable planets. These stars are presumably privileged targets for the remote spectroscopic detection of biological activity in telluric planets orbiting nearby stars by future spaceborne nulling interferometer missions such as NASA's Terrestrial Planet Finder (Lawson *et al.*, 2004), European Space Agency's Darwin DARWIN (http://ast.star.rl.ac.uk/darwin/). The detection of these so-called "biomarkers" (des Marais *et al.*, 2002, Segura *et al.*, 2003) will undoubtedly be one of the major topics of the scientific agenda of the next decades (Tarter, 2001), and represents perhaps the surest and most expedient, technologically attainable in the short term, promise of establishing with reasonable certainty the existence of life outside the Solar System.

Complex life (as we know it) is based on reasonably well established properties. The essential ones are: a long lived, solar-type star within definite constraints of luminosity; a rocky, internally hot planet able to sustain liquid water at its surface; a magnetic field able to shield the surface from biologically harmful, high-energy particles; a $CO_2$-$H_2O$-$N_2$ atmosphere with climate-regulating, carbonate-silicate cycle by way of plate tectonics (Kasting, 1996, Franck *et al.*, 1999, 2000ab, Kasting & Catling, 2003); the foundation blocks of carbon chain chemistry, water oceans and long term stability. Such definitions stem from our immediate knowledge of life on Earth, but they must surely guide our strategy to search for life elsewhere.

Stability is difficult to define. The Cretaceous-Tertiary mass extinction, 65 Myr ago, was probably triggered by a cometary impact (Alvarez *et al.*, 1980). The Permian-Triassic massive extinction 251 Myr ago, in which 96% of species vanished, bears the sign of massive volcanic eruptions and lava outflows, which boosted grenhouse gas concentrations, warming the Earth and wreaking havoc on climate, besides lowering the atmospheric $O_2$ levels to near anoxic levels (Benton & Twitchett, 2003), almost wiping out all existing life; or (Becker *et al.*, 2004) was due to yet another impact of extraterrestrial origin. There is ample evidence of past global catastrophes with heavy impact on the biota. Notwithstanding, the Earth's biosphere can and did survive these events, and thus we may take stability as, at the very least, the ability to sustain such damage for ~ 4 Gyr.

The whole issue of habitability hinges on the identification of those stars which can maintain a potentially habitable planet by providing an adequate radiation envirommnent, defining the so-called Continuously Habitable Zone (CHZ), within which liquid water can exist at planetary surfaces. The interplay of stellar radiation and climate stability seems to be fundamentally linked to the homeostatic carbonate-silicate cycle, by way of a feedback mechanism in which the atmospheric $CO_2$ concentration varies inversely with planetary surface temperature (Franck *et al.*, 2000ab, Kasting *et al.*, 1993 and references therein). This depends on the atmospheric composition and the maintenance of weathering processes by plate tectonics. Thus the inner habitability edge is defined by loss of water via photolysis and hydrogen escape, with consequent ocean evaporation. The outer edge is defined by the formation of $CO_2$ clouds which cool the planet by lowering its albedo and the atmospheric convective lapse rate. As concerns planetary mass, McElroy (1972) argues that a habitable planet must have at least a few times the mass of Mars to prevent the stripping off of its atmosphere of C, N and O atoms, by the action of the stellar wind, in timescales of a few Gyr. Kasting *et al.* (1993) also show that for planets larger than the Earth the CHZ might be slightly wider. A reasonable upper limit might be a few Earth masses, since planets larger than this may retain such a high inventory of volatiles as to be completely covered by global oceans (Ward & Brownlee, 2000, Léger *et al.*, 2004), preventing the closing of the carbonate-silicate cycle by the absence of rock weathering. Such planetary mass considerations, along with other properties of telluric planets, as well as the possibility itself of condensing them, are most probably linked to the stellar metal content (Fischer & Valenti, 2005, Santos *et al.* 2003, 2004, Lineweaver, 2001), as well as details of the planetary condensation process, and remain highly speculative under our almost total ignorance of the mechanisms responsible for telluric planet formation.

The CHZ is to be found around mid-K to late-F stars (Huang, 1959, Kasting *et al.*, 1993, Franck *et al.*, 2000a,b). Late K-type and M-type stars place their planets inside the tidal lock radius if they are to possess liquid water at the surface, whereby the planet rotates in the same time it revolves around the star, spelling potential disaster for climate stability (Kasting *et al.*, 1993, Dole, 1964). Mid-F stars have wide habitable zones but these move outward rapidly as the star evolves into subgiant status, and the time the planet can remain within the CHZ is shorter than the ~ 4 Gyr necessary for the evolution of complex life. Additionally, Catling *et al.* (2005) suggest that the partial oxygen pressure of a planetary atmosphere must surpass a threshold of ~ $10^3$-$10^4$ Pa to allow the accumulation of a sizable biomass, and that the timescale for a sufficient oxygenation is long, ~ 4 Gyr, comparable to stellar ages. This places yet another constraint against F-type stars, which might evolve too quickly, even when providing benign planetary conditions, to allow oxygenation to reach the threshold point. In mass, these limitations, for solar metallicity stars, roughly translate into the interval ~ 0.70 to 1.20 solar masses, depending, not negligibly, on the stellar metallicity, as will be discussed below. These limits can be somewhat softened, for stellar masses lower than 0.70 solar, if one accepts the possibility of a planet locking itself in orbital resonances which might preclude complete rotational synchronization, even though well before partial synchronization the planetary climate pattern would be appreciably disrupted. Masses larger than 1.20 solar may remain acceptable if allowance is made for a faster evolution of complex life than was the case on Earth. Yet, besides these well-accepted notions, quite a few other considerations have recently been increasingly considered as playing fundamental roles for the long term existence of CHZs around such stars.

The stellar Galactic orbit is a rarely mentioned factor of stability which is drawing more interest as one regards how important it may be concerning the average interval between global catastrophes that a planet may undergo. The Sun lies very near the so-called co-rotation radius (Balázs, 2000, Lépine *et al.*, 2001), where stars revolve around the Galaxy in the same period the density wave perturbations sweep across the Galactic disk. In such a situation the passages across the spiral arms, and consequently the potential encounters with star-forming regions and giant molecular clouds, are presumably minimized. The first are thought to present the danger of biologically lethal X- and gamma-ray irradiation episodes by supernova explosions (Gehrels *et al.*, 2003); the latter, to trigger heavy cometary bombardment in the inner planetary system by perturbing the Oort cloud dormant population (Clube & Napier, 1982). Leitch & Vasisht (2001) have presented some evidence on the correlation of past significant mass extinctions with crossings of the spiral arms by the Sun. Yet this issue is probably more complicated: it is conceivable that even though stars which, slightly, either lead or trail the pattern of density waves by having small velocity components with respect to them, will cross the spiral arms more often but will also remain inside the arms for a shorter time, and it remains unclear which situation presents a longer exposure to the putative hazards of this crossing. Moreover, Lépine *et al.* (2003) argue that because of resonant interaction with the spiral arms gravity fields, stellar orbits may wander as much as a few kpc from their initial position in one billion years or less. In their model, the history of spiral arm passages of any given star may be considerably subjected to random fluctuation, introducing an unforeseeable element which means that the long term survival of biospheres might be closely linked to chance. A thorough analysis of the times and frequency of stellar passages across the spiral arms, and its objective dangers, is lacking to the best of our knowledge.

Some Galactic chemical evolution considerations have also gone so far largely unappreciated as fundamental issues for providing a suitable habitable rocky planet. For its age, the Sun is quite metal rich (Rocha-Pinto & Maciel, 1996), making it one of the earliest stars of its Galactic orbit to reach metal content levels adequate to complex life. Earth might well then belong to an early generation of habitable planets in the Galaxy. Metallicity as linked to age also come to play at another point: a sizable fraction of Earth´s internal heat comes from the decay of radioactive isotopes, such as $^{40}$K, $^{235,238}$U and $^{232}$Th. Their bulk abundance in the Galaxy is synthesized by neutron capture processes in moderately or highly massive stars. Gonzalez *et al.* (2001) suggest that the relative abundances of these radioisotopes to Fe (the element taken to represent the gross stellar metal content) is decreasing with time in the Galaxy, and that future Earth-like planets will generate less radiogenic heating than the Earth does, which might be of consequence to their long term ability to maintain active plate tectonics and climate regulation. These authors further suggest, albeit rather speculatively, that there is a wide, but definite, statistical "Galactic time window" for the formation of adequate telluric planets in the Galaxy, and that past and future candidates may be less suitable to complex life than the Earth. Also, stars with lower metal content than the Sun possess a higher abundance ratio of Mg-Si to Fe, according to known trends in Galactic chemical evolution (Edvardsson *et al.*, 1993). This might also imply, for hypothetical telluric planets, a different mantle-core ratio, different liquid metallic core convection properties (and consequently a different magnetic field), and a different content of radiogenic isotopes (Gonzalez *et al.*, 2001).

A crucial parameter for the aim of the present study is the stellar age. Otherwise adequate stars may be found too close to the zero age main sequence (ZAMS), where any biosphere will necessarily be in its infancy. No complex metazoans or high $O_2$ content can reasonably be expected to have appeared, and certainly no radio-communicating civilization will exist. Alternatively, evolved stars may have outlasted their usefulness as abodes of life, if the luminosity increase sustained from zero age to their present evolutionary status surpasses the capability of the planetary thermo-regulating carbonate-silicate cycle. This situation is dependent upon planetary location inside the CHZ, yet, for excessively evolved subgiant stars, the luminosity increase may be excessive even for planets situated, initially, in the outer limits of the stellar CHZ.

Previous efforts, mainly targeted at selecting nearby stars in terms of their suitability to SETI programs, have not included the full set of criteria presented here (e.g., Blair *et al.*, 1992, Henry *et al.*, 1995), and have generally applied them with much less detail. Recently, Turnbull & Tarter (2003a,b) have applied most of the criteria presented above to the stellar sample of the *Hipparcos* catalogue (ESA, 1997), exploring its full range in distance and producing a large catalogue of stars fulfilling the criteria of luminosity class, spectral type, metallicity, age as estimated from spectroscopic indicators of chromospheric activity, and Galactic orbits inferred from velocity components. The present analysis, in contrast to theirs, is limited to those stars within 10 pc of the Sun, but is able to sieve the stars in considerably larger detail, fully exploring the completeness of data for the nearest stars and thereby being able to be less based on statistical considerations. For very nearby stars, the *Hipparcos* parallaxes allow luminosity determinations with uncertainties under 2%, and the location of their CHZ, at least for those parameters depending exclusively on the stellar luminosity, becomes very precise. These small luminosity errors also enables us to analyze each candidate star in the HR diagram appropriate to its metallicity, thus providing an independent estimate of its age, as well as an accurate evaluation of its state of evolution and increase in luminosity undergone since the stellar birth. The amount of increase in radiation any putative planet has suffered can be estimated for each case, a parameter which has been pointed out by Franck *et al.* (1999, 2000b) to have a strong bearing on the planetary biological productivity. The carbonate-silicate cycle forces the atmospheric $CO_2$ content to go steadily down as the star ages and brightens, and this essential stability mechanism eventually lowers the $CO_2$ content to a point at which photosynthesis becomes impossible even for the hardiest plants. This poses a definite upper constraint on the timescale in which even a suitable planet can sustain biological activity at its surface, and thereby maintain detectable biomarkers in its thermal spectrum, a constraint which may prove to be more stringent than the luminosity evolution of the star itself.

The full set of criteria discussed above can be used to constrain rather tightly the nearby stars in terms of suitability for complex life. The level of detail presently reachable for the nearby stars allows definite quantitative statements on habitability, and their propinquity, a key issue for the interferometric planetary detection technique, will probably make them the first interesting targets for the coming IR space-based interferometers. We aim here at characterizing which of the nearby solar-type stars can be expected to possess a reasonable CHZ and be old enough to have maintained a habitable planet within it in the last few Gyr, or at least to the point of allowing complex life and a $O_2$-rich atmosphere to have developed. It would certainly be more interesting to define complex life as a space faring culture able to establish an astrophysical understanding of its surroundings and operate radiotelescopes. However, given the failure to detect any extraterrestrial non-natural signal, despite sustained efforts (Tarter, 2001), it is probable that no such culture is broadcasting with appreciable power within our present detection limits. Thus the prospect of life detection by the photo-synthetic planetary spectral signature, with available techniques, is probably higher.

In sections 2 and 3 of this paper we detail the criteria employed to select the stars. In section 4 we discuss the results and present the best candidates, while presenting our conclusions and perspectives for future work in section 5.

# II - SAMPLE SELECTION, DATA COMPILATION AND FUNDAMENTAL CRITERIA

In this section we apply a sequence of criteria to the *Hipparcos* catalogue, removing unsuitable stars and producing a residual sample of stars which, as far as present knowledge warrants judgment, are potentially optimal hosts for habitable planets. As a first step we selected all *Hipparcos* stars within 10 pc of the Sun (figure 1), totalling 182 objects. The completeness of *Hipparcos* is 100% for V ≤ 9.0: therefore, within 10 pc, it is expected to be complete to spectral types earlier than ~ M0V, the ones relevant to our analysis. Objects closer than 10 pc which fail to make it to figure 1 owing to incompleteness of *Hipparcos* should then be exclusively late M-dwarfs[1]. The sample plotted in figure 1 is, as expected, dominated by red dwarfs.

The box plotted in figure 1 is bounded by our initial limits of color index (B-V) and absolute magnitude $M_V$ of what constitutes an astrobiologically *interesting* star, that is, a non-subgiant K2 to F8 star. These limits had to be experimented with. Without a priori knowledge of the stellar's metallicity, the colors are a poor guide to the effective temperature $T_{eff}$, since a cooler and more metal-poor star can mimic the colors of a metal-normal star: the same effect takes place for a hotter and more metal-rich star (e.g., Saxner & Hammarbäck, 1985). It will be shown below that the stellar $T_{eff}$ and metallicity govern the uncertainty in the determination of age and evolutionary mass, errors in luminosity being much less important. Initially we set as limits the values +2.0 < $M_V$ < +8.0 and +0.40 < (B-V) < +1.15, only to find that, upon an analysis in the HR diagram corresponding to the stellar metallicity, systematically too massive F-type stars appeared at the blue end of the selection, and very late-type K-stars defined the red end. Besides, such massive F-type stars, in order to be observed at all near the main-sequence, necessarily need to be very young, chromospherically active stars, and this was systematically observed. Also, in the blue, high luminosity end of this selection box, a few subgiants, too evolved to be of interest, were kept. The limits were then modified iteratively until we arrived at the following box limits: +4.0 < $M_V$ < +6.5 and +0.50 < (B-V) < +1.05. The high (B-V) limit, even for very metal-rich stars, which are redder at the same $T_{eff}$, guarantees that no late-type K-dwarf will be selected. These limits should be adequate for the metallicity range to be found in the Galactic thin disk, -0.4 < [Fe/H] < +0.4 (we use throughout the usual spectroscopic notation [A/B] = log (A/B)$_{star}$ − log (A/B)$_{Sun}$), and will be employed in a following paper in the analysis of the astrobiologically interesting stars within 10 pc and 15 pc of the Sun, for which the relevant data still lack considerably in completeness, as will be discussed below.

Including the Sun, 34 stars are kept inside the box of figure 1. We searched the literature for data on: binarity (Batten *et al.*, 1989, Duquennoy & Mayor, 1991, Warren & Hoffleit, 1987, Mason *et al.*, 2002); magneto-hydrodynamic activity, measured by the Ca II H and K lines chromospheric flux and the X-ray luminosity (Noyes *et al.*, 1984, Duncan *et al.*, 1991, Baliunas *et al.*, 1995, Henry *et al.*, 1996, Hünsch *et al.*, 1998, 1999, Strassmeier *et al.* 2000, Schmitt & Liefke, 2004); and effective temperature $T_{eff}$ and metallicity [Fe/H] (Cayrel de Strobel *et al.*, 2001, Taylor, 2003). The SIMBAD database was also scrutinized for completeness up to 2005.

All the relevant data are compiled in table 1, where we also give additional remarks on each object. The solar-type star sample comprises stars with atmospheric parameters $T_{eff}$s and [Fe/H]s spanning a fairly wide interval (figures 2 and 3). The relevant data for our discussion of the 33 solar-type stars selected by the (B-V)-$M_V$ box are given in table 1. It is noteworthy and lamentable that many of the spectroscopic analyses which provide the atmospheric parameters used in the present work leave much to be desired. Many are still based on old, low S/N photographic plates, suffering from random errors; many are not differential with respect to the Sun, suffering from systematic effects; many determine their $T_{eff}$s using only one criterium, often ill-defined photometric calibrations which do not take into account the metallicity influence on the colors; many are based on too few iron lines for a reliable estimation of [Fe/H]. It is clearly most desirable to improve upon this situation, given the modern instrumental capabilities that can be trained upon such nearby and bright stars.

The task of weeding out unsuitable stars begins by eliminating binary and multiple systems from the sample. It is still controversial whether binaries should be given lower priority as abodes of complex life. Planetary stability around binary stars is possible in definite configurations, around the system's center of mass and around each of the components, given that the planet's orbital radius is sufficiently larger than the binary separation (external case, planet around center of mass), or vice-versa (internal case, planet orbiting one of the components) (Pendleton & Black, 1993, Holmann & Wiegert, 1999, Quintana *et al.*, 2002). On the observational side, we further note that TPF mission requirements dictate a minimum acceptable binary separation, to avoid overwhelming of the planetary signal by shot noise from the secondary star (Lawson *et al.*, 2004), as 10 arcsec, and thus single stars are to be preferred.

It is clear that many planetary dynamical instabilities exist in binary or multiple systems which are absent from single stars. In the case of binaries in eccentric orbits, excitation of resonances may lead to strong instabilities and highly eccentric planetary orbits, endangering climate stability. Besides, it is a well founded expectation that the formation and evolution of protoplanetary disks, and the planetesimal condensation processes, are probably different in a binary system as compared to an isolated star, leaving at least some room to assume that binary systems are an unwanted complication. The probability of planetesimal condensation is constrained by the inclination between the accretion disk and the binary orbital plane: for large inclination angles the stability of planetary orbits is dramatically reduced (Quintana *et al.*, 2002). We further note that the very few cases of planetary systems loosely similar to our own, among the 150-plus discovered, with giant planets orbiting at > 3 AU, in approximately circular orbits, were found around single stars.

A few binary or multiple systems, however, are sufficiently interesting, by fulfilling all adequate properties but multiplicity, to warrant a full dynamical analysis. We will discuss these cases in a forthcoming paper. For our present purpose, we assume that the general constitution of the Sun's planetary system has a higher probabiblity to be assembled in single stars. We found that 13 stars, 39% of the full sample, are known astrometric and/or spectroscopic binaries, or multiple systems, and we disregard them: they are the crosses in figures 2 and 3.

The next step concerns the stellar metal abundance. Very metal poor stars, which we take here as those with approximately [Fe/H] < -0.5, are probably unable to build up a sufficiently massive telluric planet (Lineweaver 2001). No giant planet has been found so far around a solar-type star with [Fe/H] less than ~ -0.6, and the detection rate of planetary companion falls off very sharply for stars less metal rich than the Sun  (Fischer & Valenti, 2005, Santos *et al.*, 2004), at least for giant planets in orbits with periods of a few years. Constraints for the build-up of rocky planets are presumably even more severe. Therefore 4 stars, 12% of the full sample, are set apart (figure 2).

Next we eliminate the very young, chromospherically active stars from our sample. These are not necessarily inadequate to living organisms, but judging from the example of the Earth, several Gyr must pass before a biosphere with complex life allows a $O_2$-rich atmosphere to develop. Recent estimates suggest that a major increase in the oxygen content of the Earth occurred when it was ~ 3 Gyr old (Blair Hedges *et al.*, 2004), an event spurred by the advent of cell plastids and followed by the evolution of complex eukaryots.

Furthermore, near ZAMS stars have their planets, if any, in the so-called heavy bombardment period, in which planets are routinely subjected to giant meteoroid and/or comet impacts up to the 100 km mark. Even very simple life forms might find it difficult to gain a hold on planetary surfaces until the end of this phase, a few hundred Myr after the planetary system condensation. We thus concentrate our analysis in stars with at least a few Gyr of age.

In figures 2 and 3 we show, respectively, the Ca II HK chromospheric activity indicator against stellar metallicity, and the X-ray luminosity against $T_{eff}$. Data on the absolute chromospheric flux on the Ca II HK lines come from Duncan *et al.* (1991), Baliunas *et al.* (1995) and Henry *et al.* (1996): their <S> indices on the system of the Mount Wilson observatory were converted to absolute fluxes according to the recipe of Noyes *et al.* (1984). The X-ray luminosities come from Hünsch *et al.* (1998, 1999) and Schmitt & Liefke (2004). For some of the stars, data on the absolute flux in the Hα line are also available (Lyra & Porto de Mello, 2005). Chromospheric activity is a very steep function of age: particularly, the X-ray emission of young stars quickly softens after 0.5 Gyr (Güdel *et al.*, 1997), a fact of possible consequence to the evolution of the early planetary atmospheres. Young stars rapidly lose angular momentum by a magnetized wind, after ~ 1 Gyr or so, and stars older than ~ 2 Gyr tend to pile up in low activity levels (Lyra & Porto de Mello, 2005). In this fashion 10 stars are eliminated, a rather high fraction of 30% of the full sample. After the elimination of the multiple systems, metal poor and near-ZAMS stars, there remain 13 candidates: 7 are K-type, 5 are G-type, one is a F-type.

### III – ADDITIONAL CRITERIA: GALACTIC ORBITS AND ISOCHRONAL AGES AND MASSES.

The next issue concerns the stellar orbits in the Galaxy. In figure 4 we plot the mean Galactocentric radius ($R_m$) of the stellar Galactic orbits with their orbital eccentricities (Woolley *et al.*, 1970). These orbital eccentricities are not directly defined as in the two-body problem, the actual orbits being open, but are defined as e = $(R_a-R_p)/(R_a+R_p)$, $R_a$ and $R_p$ being the maximum (apogalacticum) and minimum (perigalacticum) distances from the Galactic center: the mean orbital distance is simply defined as $R_m = (R_a+R_p)/2$. The values of Woolley *et al.* (1970) were calculated with a rather simple model of the Galactic potential, and were based on old pre-*Hipparcos* parallaxes: for these very nearby stars, however, the new parallaxes differ very little from those used by Woolley *et al.* (1970). Besides, the mean distances $R_m$ and orbital eccentricities calculated by Woolley *et al.* (1970) differ from those of Edvardsson *et al.* (1993), obtained from a more up-to-date code for the Galactic potential, by ~ 0.4 kpc in $R_m$ and ~ 0.04 in eccentricity, in average, for six stars in common: these differences have little bearing on our conclusions.

We have also calculated the UVW components of the stellar space velocities (Johnson & Soderblom, 1987), relative to the Sun, from parallaxes and proper motions from *Hipparcos*, and radial velocities from the *SIMBAD* database. The components are plotted in figures 5 and 6, in the usual notation: U, V, and W are positive in the direction of the Galactic centre, Galactic rotation, and north Galactic pole, respectively. In these graphs, the 1σ and 2σ ellipsoids of the dispersions of the velocity components, with respect to the Sun, for stars with [Fe/H] > -0.4 in the Edvardsson *et al.* (1993) sample of thin disk stars, as calculated by Turnbull & Tarter (2003a), are also shown. The kinematical parameters are given for all 33 stars plus the Sun in table 2. It can be seen that the range of the kinematical parameters is considerable.

In figure 4 the well-known "wedge" effect is apparent: stars with Galactocentric distances far from the Sun's systematically display high eccentricity orbits. This is a necessary condition for them to be observed at all in the solar neighborhood. It is striking in figure 4 that the Sun possesses a very small orbital eccentricity, e = 0.06, which, along with its mean $R_m$ near the co-rotation radius, makes for a near circular, stable orbit, and a small number of spiral arm passages. A large fraction of the local stars have orbital eccentricities higher than solar, which makes them but temporary dwellers of the solar vicinity. Their space velocities eventually carry them either far away to the outskirts of the Galactic disk, where they are overtaken by the spiral arm density wave, or deeper into the inner disk, where they themselves overtake the waves. Both situations subject them to more encounters with star-forming regions then the Sun. Albeit with large uncertainty, the number of spiral arm encounters that a star experiences in its life can be calculated, but no comprehensive analysis on this, for the nearby population of solar-type stars, has been published to the best of our knowledge; neither has a systematic, volume-limited and up-to-date review of the mean Galactocentric distances and orbital eccentricities of nearby stars. Nevertheless, quantitative discussions of the hazards of spiral arm crossing remain highly speculative to this date. Therefore we are not expelling stars from our final best candidate list on the basis on the Galactic orbit parameters. In figure 4 we merely show, inside the box, the three stars with orbital parameters truly similar to the Sun's, for which a similar history of passages through the Galactic spiral arms could be reasonably expected. Another group of nine stars have orbital eccentricities within ~ 0.15-0.20, and their Galactic orbits have probably allowed a larger number of spiral arm passages than the stars with eccentricities less than ~ 0.10. One star, HD 185144, has such a large orbital eccentricity as to have very probably experienced a totally different history of interaction with the spiral arms than has been the case for the Sun.

In figures 5 and 6 it is seen that few stars fall significantly outside the $2\sigma$ ellipsoids either in the UV-plane or the VW-plane. The UV-plane is the most relevant to our discussion, since the U and V velocities are the one which determine the orbital eccentricity and mean Galactocentric distance in the plane of the Galactic disk, and thus the number of spiral arm passages undergone. The W component actually determines the amount of time the stars spends vertically away from the Galactic disk, as well as the maximum distance reached above the disk. Larger values of this component may merely mean that, unfavorable U and V components notwithstanding, the star will tend to experience less passages inside the disk and less encounters with spiral arms. In the UV-plane, the outliers are HD 4628, 16160, 32147, 102365, 115617 and 185144, though none of them significantly. As will be seen below, all of these are interesting candidates after consideration of all criteria. In the VW-plane, HD 1581, 4628, 32147 and 115617 are outliers. Three stars, HD 4628, 32147 and 115617, have all component velocities rather distinct from the Sun's. It is also seen that only few stars, in figures 5 and 6, possess velocity components within $1\sigma$ of the solar ones or nearly so, HD 10476, 109358 and 190248. The last two also very closely share the galactic orbits of the Sun in figure 4.

Now we examine each star in detail in the theoretical HR diagrams of Kim *et al.* (2002) and Yi *et al.* (2003), in order to assess their masses and isochronal ages. It is known that chromospheric activity indicators are very sensitive to age for the first ~ 2 Gyr of a star's life (Lyra & Porto de Mello, 2005, Güdel *et al.*, 1997), but lose considerably this sensitivity as stars age, so that beyond the solar age, ~ 4.6 Gyr, virtually no age-ability discrimination remains. At this point, for stars older than the solar age, theoretical isochrones possess good age discrimination, provided that the stellar $T_{eff}$ and metallicity are know with sufficient accuracy and the stars are not too close to the ZAMS. Each diagram has been calculated, using routines provided by the authors, for a set of stellar masses ranging from 0.6 to 1.3 solar masses, corresponding to a given metallicity. Taking into account, conservatively, the uncertainty of the $T_{eff}$s and metallicities of our sample, both from the heterogeneity of sources and the errors inherent to spectroscopic analyses, we have grouped the candidate stars in intervals of $\pm0.1$ dex around a value [Fe/H] chosen such that each star is evaluated in the HR diagram corresponding to a metallicity very similar to its own. For the six stars with more than three spectroscopic determinations of $T_{eff}$ and [Fe/H], we determined the standard deviations around the mean of these quantities considering all published values as, respectively, 70 K and 0.08 dex. The results are shown in figures 7 to 13, where we adopt 100 K and 0.1 dex, rather conservatively, as estimate of the respective uncertainties in $T_{eff}$, and [Fe/H].

We note that the results of the HR diagram analysis are weakly dependent on the $T_{eff}$ and [Fe/H] errors, as long as the goal remains at identifying stars which are older than a few Gyr: better age discrimination would require much more accurate $T_{eff}$s. Some allowance also has to be made to uncertainties in the models, even though the Sun state of evolution is well reproduced for the correct age (figure 12).

The CHZ positions essentially depend only on the stellar luminosity, known to better than 2% in all cases studied here: effects on these positions of varying the stellar $T_{eff}$s are second order (Kasting *et al.*, 1993). The uncertainty in determining the CHZ locations thus come almost exclusively from the planetary climate theory. Stellar ages inferred from HR diagrams do depend rather heavily on $T_{eff}$, but they are also constrained by the chromospheric activity indicators. The conservative allowance of 100 K we made for the $T_{eff}$ errors, even though affecting the age estimates, do not change any of the conclusions presented.

## IV – DISCUSSION

The 13 stars which survived all our criteria range somewhat in mass, age, metallicity and orbital properties: a qualitative assessment of these is given in table 3, where stars are rated according to their similarity with the solar parameters. According to the models of Kasting *et al.* (1993), there are definite luminosity limits for the succesful operation of the planetary climatic feedback mechanisms: these limits are given as three cases. In the so called "pessimistic" scenario, the feedback is considered to fail, at the "hot" end, with the onset of ocean evaporation, beginning at a luminosity of $L/L_\odot \sim 1.10$; at the "cool" end, with the formation of $CO_2$ clouds, beginning at $L/L_\odot \sim 0.53$. In a somewhat less stringent view, failure happens at the "hot" end only with the runaway greenhouse, by which water loss happens in a short timescale as compared to geologic time: this corresponds to $L/L_\odot \sim 1.41$. At the "cool" end, the planet cannot be further warmed when the maximum-greenhouse limit is reached, at which the partial pressure of $CO_2$ is so high that clouds form and increase the planetary albedo, at a luminosity of $L/L_\odot \sim 0.36$. A very optimistic possibility discussed by Kasting *et al.* (1993) is the case where Venus is considered to have had a water ocean for as long as a few Gyr, and Mars is considered to have been sufficiently warm in its early evolution to have large standing bodies of water: the corresponding luminosity limits, respectively, would be $L/L_\odot \sim 1.76$ and $L/L_\odot \sim 0.32$ in this case. Considerable uncertainty still affects the luminosity values for each of the above cases.

For the present discussion, the important issue is that there is a maximum luminosity increase that a star may undergo, once a planet is initially placed in the most favorable configuration for its maximum stay at the CHZ, its outer limit, before the climatic feedback stops working. This maximum luminosity increase for the "pessimistic" case, the one we shall adopt here, would thus be $L_{incr} \sim 2$, being slightly mass dependent; it is $L_{incr} \sim 4$ for the less rigid case.

In figures 7 to 13, an evaluation of the evolutionary state and age of the stars can be made whenever the data allows a good match with theory. Metallicity is a key quantity: its accurate determination enables a correct choice of the set of evolutionary diagrams to be used in the determination of the stellar mass and age. Effective temperature more directly affects the mass and age determinations. Metallicity also has an important bearing on the evolutionary timescale. At the same mass, metal poor stars have less opacity and therefore have their radiative energy blanketed farther away from the core. This makes them more luminous and their rate of evolution faster, and it takes them less time to grow too luminous for a given CHZ location. The difference is not inconsiderate: for a solar mass star, and evolutionary times comparable to the solar age, the time spent to reach a given increase in luminosity is ~ 20% less for a star with half the solar metal content. The corresponding time is ~ 30% longer for a star with double the solar metal content. This evolutionary timescale can therefore be more than 50% longer for a metal rich star, as compared to a metal poor star, for the same mass, considerably affecting the lifetime of planets within the stellar CHZs.

We now discuss each case: figure 7 shows HD 1581 to be a very good candidate. Its mass and metallicity are very close to solar. Its position in the HR diagram might be compatible with a near-ZAMS status judging by the $T_{eff}$ error bar, but the chromospheric activity data consistently point towards an age greater than 3 Gyr – it could be compatible with the solar age for a slightly larger mass.

The data therefore place it as very probably slightly more massive and slightly younger than the Sun, and it has probably also suffered less evolutionary change..

In figure 8 three candidates with [Fe/H] = -0.27 are plotted. HD 102365 turns out to be an interesting candidate, being slightly less massive than the Sun, but considerably older: its age can be confidently determined as larger than ~ 7 Gyr. It has suffered a luminosity increase, since the ZAMS phase, $L_{incr}$ ~ 1.9, approaching the limits for the "pessimistic" case, and this fact, along with an appreciably lower metallicity, would suggest a not so high priority. HD 100623 is seen not to be very well fitted by the theoretical tracks, but it is barely compatible with a mass of 0.80 $M_\odot$ and an age ~ 7 Gyr. HD 4628 is not fitted at all by the models, lying well above the tracks for any reasonable age, and its age cannot be constrained. Lebreton *et al.* (1999) discuss the incompatibility of theoretical models and the stellar positions in the HR diagram for moderately metal-poor stars, [Fe/H] < -0.45: these stars are displaced towards lower $T_{eff}$s than the models predict, exactly the effect seen in figure 8. This discrepancy is attributed by Lebreton *et al.* (1999) to deviations from LTE and the sedimentation of helium and heavy metals towards the stellar interiors in timescales approaching 10 Gyr. In their comparisons, the effect seems to be slightly enhanced for the lower mass stars. Taking into account the composed effect of non-LTE corrections and the diffusion of metals improves considerably the agreement between theory and observations. Another possible source of the inability of the models of Kim *et al.* (2002) and Yi *et al.* (2003) to reproduce the low mass, low metallicity stars is the inherent difficulty in determining the $T_{eff}$ of these objects. The differential technique is normally applied in the spectroscopic analyses of solar-type stars, and the greater the difference between the stellar and solar $T_{eff}$ the higher possibility of systematic errors, with direct bearing on the [Fe/H] determination. Thus, the results for the K-stars must be regarded as less reliable. In the present case, HD 100623 could be considered as marginally compatible with an age higher than the Sun, while HD 4628 is unconstrained in age, besides having a probable mass close to the lower limit for the planet to remain tidally unlocked inside the CHZ.

In figure 9 we consider the slightly metal-poor star HD 185144: it is reasonably compatible with an age greater than ~ 8 Gyr and a mass of 0.80 $M_\odot$, but assigning lower masses to it would lead to unreasonable ages, greater than the age of the Galactic disk, ~ 8-10 Gyr (del Peloso *et al.* 2005b). In figure 10, HD 32147 is seen to be another case of a K-dwarf with totally unconstrained age: its $T_{eff}$ error bar allows any age value from the ZAMS to the age of the Galaxy, whilst its mass lies probably in the 0.80-0.85 $M_\odot$ range. HD 32147 has a rather high metallicity of [Fe/H] = +0.34, and non-LTE effects are not a reasonable explanation for the discrepancy (Lebreton *et al.* 1999): a very old age and metal diffusion could be the culprits here.

Figures 11 and 12 present two excellent candidates. HD 109358 is slightly metal-poor, its mass is strictly within the solar range, its age is well constrained in the 3-8 Gyr range and its luminosity increase since the ZAMS phase has been $L_{incr}$ ~ 1.5, similarly to the Sun, $L_{incr}$ (Sun) ~ 1.4. HD 115617 is plotted along with the Sun, HD 10476 and HD16160 in a diagram corresponding to the solar metallicity: its age range lies within the 3-11 Gyr interval, and it is compatible with a mass slightly lower than the Sun, and the same age, although with large uncertainty. HD 10476 is essentially unconstrained in age, but could be compatible with ~ 6 Gyr, and a mass slightly larger than 0.85 $M_\odot$. If it is indeed an old star, heavy element diffusion could explain its discrepant position in the HR diagram. HD16160 lies below the ZAMS, but its $T_{eff}$ error bar places it within compatibility with a mass in the 0.80-0.85 $M_\odot$ range, even though its age cannot be constrained in the diagram. This solution would place it at the ZAMS, and we note that its chromospheric activity indicators in the literature are inconsistent (table 1), some suggesting low activity and old age and others the opposite, leaving the status of this object uncertain.

Figure 13 presents two low mass candidates, HD 192310 and 219134. HD192310 is barely compatible, at the hot end of its $T_{eff}$ uncertainty, with a mass of ~ 0.85 $M_\odot$ and a very old age. Lower masses are untenable since they would lead to unreasonably old ages: its discrepant position in the HR diagram could again be explained by metal diffusion and old age. HD 219134 lies below the ZAMS but could be reconciled with a mass of ~ 0.80 $M_\odot$ if its $T_{eff}$ were cooler by ~ 150 K. Yet even then it would be placed at the ZAMS, and all chromospheric activity data discredit this option: it is less active than HD192310, already an unactive star.

The last candidate, HD 190248, is shown in figure 14. It is the nearest of the 13 final candidates, and a super-metal rich star (del Peloso *et al.*, 2005a), appreciably more luminous than the Sun, and certainly an evolved star, in the process of becoming a subgiant and close to the turn-off point: its age is well constrained, in the 5-7 Gyr interval, by its position in the locus of the evolutionary tracks where they are parallel to the $T_{eff}$ axis, even though it is close to the region in the HR diagram where the isochrones overlap and an unambigous age determination becomes impossible. Its luminosity has increased ~ 60% since zero age, similarly to the Sun. It is a fair example of slow stellar evolution owed to a high metallicity: at its derived mass of 1.25 $M_\odot$, a solar metallicity star aged 5 Gyr would be far up the red giant branch, having reached a luminosity 40 times greater then solar. High metallicity stars more massive than the Sun may thus become interesting astrobiological targets owing to their slow evolution, besides wide CHZs due to high luminosity. HD 190248 is certainly more massive and cooler than the Sun, and actually a rather interesting object: its high metallicity probably lies at the high tail of the thin disk distribution (Castro *et al.*, 1997, Bodaghee *et al.*, 2003). Its old age, coupled to its high metallicity, implies that it is a fairly non-typical object for its epoch of formation and Galactic orbit, which closely matches the Sun's (figure 4). Turnbull & Tarter (2003b) consider it the nearest interesting SETI target. Santos *et al.* (2004) find that at the metallicity of HD 190248, 30% of nearby stars possess at least one giant planet, most of them in close-in orbits, yet none has been reported so far for HD 190248. Not one of our 13 final candidates have had any detection of planetary companions as we write.

We consider as the optimum candidates those stars with mass and metallicity close to the solar ones, besides an age compatible with our working hypothesis of a minimum of ~ 3 Gyr elapsed before putative habitable planets can develop a sufficiently high atmospheric $O_2$ content to be detectable remotely. The last criterium is a stellar orbit reasonably similar to the Sun's. In table 3 we qualitatively rank the stars as candidates comparing these four parameters to the solar one, giving one asterisk in the final column for each parameter similar to the solar one.

One of the candidate stars, HD 102365, has undergone a greater increase in luminosity since the ZAMS phase than the Sun, and is on its way to become a subgiant. Present subgiants may have been pleasant hosts to planets suitable to complex life during their stay in the main-sequence, but as they evolve the planet is hard put to remain inside the CHZ. The Earth will eventually be rendered uninhabitable by the increased solar luminosity: at a luminosity of $L/L_\odot$ ~ 1.1, water loss begins. In figure 12 it is seen that the models of Kim *et al.* (2002) and Yi *et al.* (2003) place this event approximately 1 Gyr in the future. However, Franck *et al.* (1999) model the evolution of the $CO_2$ content in the Earth's atmosphere, coupled to the solar radiation, the variation of continental area, the weathering rate and the bioproductivity of the biosphere (biomass per area per time), under the influence of the external forcing of the Sun's luminosity evolution. They conclude that the lowering of the $CO_2$ partial pressure, necessary to counter the forcing of the increasing solar radiation, will render photosynthesis impossible ~ 0.6 to 0.8 Gyr from now, the results being not very sensitive to the adopted model of continental area evolution. It seems possible that such issues ultimately govern a planet's stay inside the CHZ. Therefore, stars close to the maximum luminosity increase adopted by us for the "pessimistic" case, ~ 2, are less promising candidates.

A possible drawback of 6 of the final 13 candidates (table 3) is their moderately low metallicity, ~ 50-60% solar, which might carry a lower probability of formation of telluric planets, although recent results imply a constant frequency of giant planets in stars within the range of metallicity from solar to one fourth solar (Santos *et al.*, 2004). Such lower metallicity might also imply, for hypothetical telluric planets, a higher abundance ratio of Mg-Si to Fe, according to known trends in Galactic chemical evolution (Edvardsson *et al.*, 1993), a different mantle-core ratio, different liquid metallic core convection properties (and consequently a different magnetic field), and a different content of radiogenic isotopes (Gonzalez *et al.*, 2001), all these factors with presumable bearing on the habitability of telluric planets

## V – CONCLUSIONS

After all criteria are considered, we end up with 13 good candidates for nearby "biostars", a fraction of roughly 40% of the nearby solar-type stars, and 7% of all stars within 10 parsecs (excluding a possible incompleteness for the latest M-dwarfs). Three of them, HD 1581 (ζ Tucanae), HD 109358 (β Comae Berenicis) and HD 115617 (61 Virginis) rank especially high in that they have properties reasonably similar to the Sun's, concerning mass, age, metallicity and evolutionary status. If we relax somewhat the orbital eccentricity criterium and allow for the fact that HD 1581 and 115617 have orbital eccentricities e ~ 0.15, they would rank well in all criteria. We suggest these objects as high priority targets in SETI surveys and in future space interferometric missions aiming at detecting life by the ozone atmospheric infrared biosignature of telluric planets.

The basic astrophysical information on the nearby solar-type stars is fairly complete, although, in many cases, still innacurate and/or of low-quality. Yet, important details are still lacking. Chromospheric activity indicators provide a broader picture that constrain stellar ages. However, it is well known that different chromospheric flux diagnostics may lead to different conclusions. For example, it has been suggested that the Sun is less active and shows a longer cycle (Hall & Lockwood, 2001) than its almost perfect twin HR6060 (Porto de Mello & da Silva, 1997) as judged by its Ca II HK flux. However, when evaluated by its Hα flux, the Sun does not seem to be an exceptional star (Lyra & Porto de Mello, 2005). Isochronal ages for moderately evolved stars may be well constrained in theoretical HR diagrams if metallicities and effective temperatures are known with high accuracy. These call for detailed, high-quality, homogeneous spectral analyses. It would also be very interesting to study the abundances of elements other than Fe in the photospheres of the "biostars".

We fully anticipate that the astrobiological suitability criteria here discussed will have to be revised as our knowledge deepens. Important present themes of research include more detailed models of the planetary climate stability and a possible relaxation of the exclusion of late-K and early M-type stars as inadequate, as well a clearer assessment of the importance of stellar metallicity for the formation of planets. Also, an analysis if the effective biological dangers of Galactic orbits which subject stars to frequent crossings of spiral arms would be of great interest.

An observational program to extend the present analysis to greater distances from the Sun, and to determine homogeneously, with high quality data, $T_{eff}$s, chemical composition, chromospheric activity level, evolutionary status, mass and age of the astrobiologically interesting stars of the solar neighborhood, is under way.

### ACKNOWLEDGEMENTS

G.F.P.M. acknowledges financial support by FAPESP/Projeto Temático (grant 00/06769-4), CNPq/Conteúdos Digitais (grant 552331/01-5) and CNPq/MEGALIT/Institutos do Milênio program, and a FAPERJ (grant APQ1/26/170.687/2004). E.F.d.P. thanks FAPERJ for a post-doctorate grant (E-26/150.567/2003) and CNPq for a DTI grant (382814/2004-5). L.G. thanks CNPq/Brazil for a PIBIC grant. This research has made use of the SIMBAD database, operated at the CDS, Strasbourg, France.

TABLE 1 – Essential data for the 33 solar-type stars within 10 parsecs of the Sun which are candidates to astrobiologically interesting stars: the Sun is included as the first entry. The first four columns are respectively the HD number, name, distance in parsecs and spectral type; the fifth and sixth columns are the effective temperature (in Kelvin) and metallicity; the seventh column is the logarithm of the stellar luminosity with respect to the Sun; the eighth and ninth columns are the X-ray luminosities, in $10^{27}$ erg s$^{-1}$, and the logarithm of the chromospheric flux in the Ca II H and K lines, in erg cm$^{-2}$ s$^{-1}$ (see text for sources); the tenth column provides remarks under which each object was disregarded as an astrobiologically interesting target by our criteria, and the eleventh column provides the source of metallicity and effective temperature. Variability data from *Hipparcos* are given in the remarks to the table.

| HD | Name | d (pc) | Type | $T_{eff}$ (K) | [Fe/H] | log L/L$_\odot$ | L$_X$ | log R'$_{HK}$ (erg s$^{-1}$) | Remarks | Source |
|---|---|---|---|---|---|---|---|---|---|---|
| --- | Sun | --- | G2V | 5777 | +0.00 | +0.00 | 3.8 | -4.85 | --- | --- |
| 1581 | ζ Tuc | 8.6 | F9V | 5970 | -0.07 | +0.07 | --- | -4.83 | | Castro *et al.* 1999 |
| 4614 | η Cas | 6.0 | K2V | 5900 | -0.25 | +0.07 | 2.6 | -4.94 | binary | Cunha *et al.* 1995 |
| 4628 | --- | 7.5 | G0V | 4910 | -0.27 | -0.47 | 0.9 | -4.88 | | Zboril & Byrne 1998 |
| 6582 | μ Cas | 7.6 | G5V | 5390 | -0.83 | -0.29 | --- | -4.92 | binary, metal poor | Zhao & Gehren 2000 |
| 10360 | --- | 8.1 | K0V | 5045 | -0.19 | -0.40 | --- | -4.85 | binary | Santos *et al.* 2001 |
| 10476 | 107 Psc | 7.5 | K1V | 5200 | +0,03 | -0.30 | --- | -4.87 | | Heiter & Luck 2003 |
| 10700 | τ Cet | 3.6 | G8V | 5320 | -0.50 | -0.26 | 0.2 | -4.91 | metal poor | Castro *et al.* 1999 |
| 10780 | --- | 10.0 | K0V | 5350 | +0.07 | -0.23 | 22 | -4.60 | binary, near ZAMS | Heiter & Luck 2003 |
| 16160 | --- | 7.2 | K3V | 5100 | -0.03 | -0.52 | 1.7 | -4.94 | | Heiter & Luck 2003 |
| 20630 | κ Cet | 9.2 | G5V | 5600 | +0.04 | -0.07 | 77.1 | -4.42 | near ZAMS | Cayrel de Strobel & Bentolila 1989 |
| 20794 | 82 Eri | 6.1 | G8V | 5360 | -0.54 | -0.13 | 0.5 | -4.93 | metal poor | François 1986 |
| 22049 | ε Eri | 3.2 | K2V | 5180 | -0.09 | -0.40 | 20.7 | -4.46 | binary, near ZAMS | Drake & Smith 1993 |
| 26965 | o$^2$ Eri | 5.0 | K1V | 5185 | -0.26 | -0.32 | 4.1 | -4.85 | triple | Bodaghee *et al.* 2003 |
| 32147 | --- | 8.8 | K3V | 4825 | +0.34 | -0.50 | 1.5 | -4.99 | | Thoren & Feltzing 1998 |
| 39587 | χ1 Ori | 8.7 | G0V | 5929 | -0.02 | +0.01 | 120 | -4.46 | binary, near ZAMS | Castro *et al.* 1999 |
| 100623 | --- | 9.5 | K0V | 5400 | -0.26 | -0.38 | --- | -4.86 | | Flynn & Morell 1997 |
| 101501 | 61 UMa | 9.5 | G8V | 5600 | +0.00 | -0.18 | 16.2 | -4.50 | near ZAMS | Heiter & Luck 2003 |
| 102365 | --- | 9.2 | G5V | 5643 | -0.28 | -0.07 | --- | -4.90 | | Porto de Mello 1996 |
| 109358 | β CVn | 8.4 | G0V | 5860 | -0.21 | +0.06 | 0.6 | -4.85 | | Fuhrmann *et al.* 1998 |
| 114710 | β Com | 9.2 | G0V | 5952 | +0.00 | +0.12 | 11.5 | -4.75 | near ZAMS | Gratton et al. 1996 |
| 115617 | 61 Vir | 8.5 | G5V | 5587 | +0.00 | -0.07 | --- | -4.96 | | Porto de Mello 1996 |
| 128620 | α Cen A | 1.3 | G2V | 5857 | +0.23 | +0.21 | 2.2 | -4.76 | binary | del Peloso *et al.* 2005a |
| 128621 | α Cen B | 1.3 | K1V | 5300 | +0.20 | -0.22 | 2.2 | -5.16 | binary | Chmielewski *et al.* 1992 |
| 131156 | ξ Boo | 6.7 | G8V | 5500 | -0.15 | -0.16 | 87.3 | -4.32 | binary, near ZAMS | Ruck & Smith 1995 |
| 149661 | 12 Oph | 9.8 | K2V | 5300 | +0.01 | -0.29 | 14.5 | -4.55 | near ZAMS | Flynn & Morell 1997 |
| 155885 | 36 Oph B | 6.8 | K1V | 5140 | -0.35 | -0.12 | 19.2 | -4.50 | triple, near ZAMS | Cayrel de Strobel *et al.* 1989 |
| 156274 | 41 Ara | 8.8 | G8V | 5305 | -0.35 | -0.27 | 1.9 | -4.95 | binary | Perrin *et al.* 1988 |
| 165341 | 70 Oph | 5.1 | K0V | 5260 | -0.25 | -0.13 | 28.3 | -4.57 | binary, near ZAMS | Zboril & Byrne 1998 |
| 185144 | σ Dra | 5.8 | K0V | 5310 | -0.21 | -0.31 | 4.1 | -4.81 | | Clegg *et al.* 1981 |
| 190248 | δ Pav | 6.1 | G7IV | 5588 | +0.38 | +0.11 | 1.8 | -4.97 | | del Peloso *et al.* 2005a |
| 191408 | --- | 6.1 | K2V | 4890 | -0.58 | -0.48 | --- | -5.04 | metal poor | Abia *et al.* 1988 |
| 192310 | --- | 8.8 | K3V | 5125 | +0.05 | -0.36 | 1.7 | --- | | Bodaghee *et al.* 2003 |
| 219134 | --- | 6.5 | K3V | 5100 | +0,10 | -0.47 | 0.4 | -5.08 | | Heiter & Luck 2003 |

Remarks to table 1 (HK refers to the chromospheric Ca II H & K line flux, see sources in text; Hα to the chromospheric flux data from Lyra & Porto de Mello 2005):

HD 1581: Low average HK flux and a quiet Ca II K line profile (Pasquini 1992). The Hα flux is low and also compatible with an age of a few Gyr. It is listed among the least variable stars in the *Hipparcos* catalogue (Adelman 2001). Endl *et al.* (2002) find its radial velocity as constant within ~ 22 m s$^{-1}$, ruling out "hot jupiters" for this object.

HD 4614: Binary, period 480 years, component A more evolved than the Sun, masses 0.95 and 0.62 (Fernandes *et al.* 1998) solar masses. Low L$_X$ and HK flux: age is 3 to 4 Gyr (Fernandes *et al.* 1998). *Hipparcos* data list separation 12.49" and magnitude difference Δm = 3.78. Component A has been reported to be a spectroscopic binary, but this is not confirmed. *Hipparcos* data report unsolved photometric variability.

HD 4628: Monitoring of its HK chromospheric flux (Duncan *et al.* 1991) indicate a uniformly low activity level, in excellent compatibility with a low L$_X$ and long rotational period.

HD 6582: Spectroscopy binary, period 22 years, orbital eccentricity 0.58 (Russell & Gatewood 1984). Low HK flux.

HD 10360: Binary, P = 484 years, pair of very similar K0V dwarfs. *Hipparcos* data list separation 11.26" and magnitude difference Δm = 0.00. Low HK flux. *Hipparcos* data report duplicity induced variability.

HD 10476: Listed as a binary, unresolved by speckle interferometry (Hartkopf & McAlister 1984), but given as single in the Washington Double Star catalog (WDS, Mason *et al.* 2002) and by HK flux lower than solar.

HD 10700: Chromospheric very low HK flux and low Hα flux combine to suggest an age appreciably larger than solar.

HD 10780: High L$_X$ and HK flux, very young star.

HD16160: Very low $L_X$ and HK flux lower than solar, yet Cutispoto et al. (2002) detect HK emission with a central reversal, in their analysis of fast-rotating stars: strong evidence of chromospheric activity and youth. Endl *et al.* (2002) find its radial velocity constant within 10 m s$^{-1}$, ruling out "hot jupiters".

HD 20630: Very high $L_X$ and HK flux, short rotational period of 9.4 days; a high H$\alpha$ flux is compatible with an age ~ 1 Gyr. The WDS gives component A as a spectroscopic binary. *Hipparcos* data report a photometric period of 9.09 days, in good agreement with the rotational period.

HD 20794: Very low $L_X$ and HK flux.

HD 22049: Extremely high HK flux, high $L_X$, rotational period of 11 days, H$\alpha$ is compatible with an age of a few hundred million years, a very young star. Astrometric binary with a period of 25 years according to the Bright Star Catalogue (Warren & Hoffleit 1987), unconfirmed by Batten *et al.* (1989). Planetary companion detected (Hatzes *et al.* 2000) with a period of 2502 days, minimum mass 0.86 Jupiter masses, semi-major axis of 3.3 AU, orbital eccentricity 0.60. Possible microvariable (*Hipparcos*).

HD 26965: Triple system: component B is a white dwarf, component C is a red dwarf with strong X-ray emission. Component A has $L_X$ and HK flux similar to solar. Unsolved variable (*Hipparcos*).

HD 32147: Very low $L_X$ and HK flux.

HD 39587: Binary with a M-dwarf, P = 14.2 years, mass of secondary approximately 0.15 solar masses (Irwin *et al.* 1992), member of the 0.3 Gyr old Ursa Major moving group (Castro *et al.* 1999). Very high activity level in X-ray, HK and H$\alpha$, rotational period of 5.4 days: probably a very young star. Possible microvariable (*Hipparcos*).

HD 100623: Activity level similar to the Sun's.

HD 101501: High $L_X$ and HK flux, rotational period of 17 days, young star. Unsolved variable (*Hipparcos*).

HD 102365: Very low HK flux, H$\alpha$ flux much lower than solar. Endl *et al.* (2002) find its radial velocity as constant within ~ 15 m.s$^{-1}$, ruling out "hot jupiters" for this object.

HD 109358: HK flux and $L_X$ lower than solar. The WDS lists a companion 0.1" away, citing data probably from McAlister (1978), and probably in error (Turnbull 2005), since this author gives a negative result for HD 109358. Batten *et al.* (1989) give it single status. The WDS data probably refer to the first suggestion by Abt & Levy (1976) that this star might have a spectroscopic companion: this suggestion was convincingly refuted by Morbey & Griffin (1987).

HD 114710: High $L_X$, HK flux suggests the same level of activity as the Sun or slightly higher, H$\alpha$ flux slightly lower than solar, rotational period 12 days. This data together with a high lithium abundance of log N(Li) = 2.60 (Mallik 1998) place it very probably as appreciably younger than the Sun. The WDS considers it a possible long period spectroscopic binary.

HD 115617: H$\alpha$ slightly higher than solar, HK lower than solar.

HD 128620: $\alpha$ Centauri A, astrometric binary, P = 81.2 years, masses 1.17 (A) and 1.09 (B) solar masses (Pourbaix *et al.* 1999), H$\alpha$ flux slightly higher than solar, HK flux and $L_X$ lower than solar. Duplicity induced variability (*Hipparcos*).

HD 128621: $\alpha$ Centauri B (see above), H$\alpha$ flux higher than solar, HK flux and $L_X$ lower than solar. Duplicity induced variability (*Hipparcos*).

HD 131156: Binary, P = 151 years, masses 0.89 (A) and 0.79 (B) solar masses, very high $L_X$, HK and H$\alpha$ flux for spectra of combined components, rotational periods 6.2 (A) and 11.5 (B) days. *Hipparcos* data list separation 7.07" and magnitude difference $\Delta$m = 2.27, and reports unsolved variability.

HD 149661: High $L_X$, high HK flux, rotational period 21.3 days: young star.

HD 155885: Member of triple system of K dwarfs, P = 549 years, rotational period 22.9 days (A), high $L_X$, high HK flux: young star. *Hipparcos* data list separation 4.74" and magnitude difference $\Delta$m = 0.01. *Hipparcos* data report duplicity induced variability.

HD 156274: Binary, companion M-dwarf, published periods range from 693 to 2205 years. H$\alpha$ flux equal to the Sun's, very inactive star in $L_X$ and HK. *Hipparcos* data list separation 8.66" and magnitude difference $\Delta$m = 3.22, and reports unsolved variability.

HD 165341: Spectroscopic binary, two K dwarfs, P = 88 years, eccentricity 0.50. Very high activity level, rotational period 19.7 days (A). *Hipparcos* data list separation 1.59" and magnitude difference $\Delta$m = 1.84, and reports periodic photometric variability at 1.96 days.

HD 185144: Low $L_X$, low HK flux. It is listed among the least variable stars in the *Hipparcos* database.

HD 190248: Low $L_X$, low HK flux, H$\alpha$ flux slightly higher than solar.

HD 191408: Common proper motion with M dwarf; Endl *et al.* (2002) find no radial velocity variation compatible with stellar or substellar companions. Low HK flux, but H$\alpha$ flux higher than solar suggests age of less than 1 Gyr, which is corroborated by detection of infrared excess compatible with Vega-like disk (Laureijs *et al.* 2002). Probably a young star.

HD 192310: Low $L_X$, flux in He I D3 line lower than solar (Saar *et al.* 1997).

HD 219134: Low $L_X$, very low HK flux.

TABLE 2 – Space motion components, with respect to the Sun, and Galactic orbit parameters for the 33 solar-type stars within 10 parsecs of the Sun which are candidates to astrobiologically interesting stars: the Sun is included as the first entry. Columns 2, 3 and 4 give the velocity components, in km s$^{-1}$, respectively, positive in the direction of the Galactic center, the Galactic rotation and the Galactic north pole. Proper motions and distances were taken from the *Hipparcos* catalogue; radial velocities from the SIMBAD database. Columns 5 and 6 give the orbital eccentricity and the mean Galactocentric distance, taken as the mean of the apogalactic and perigalactic distance, (Woolley *et al.* 1970), normalized to a solar Galactocentric distance of 8.5 kpc.

| HD | U (km s$^{-1}$) | V (km s$^{-1}$) | W (km s$^{-1}$) | ecc. | $R_m$ (kpc) |
|---|---|---|---|---|---|
| Sun | 0 | 0 | 0 | 0.06 | 8.5 |
| 1581 | -38 | +16 | -39 | 0.16 | 8.4 |
| 4614 | -18 | -2 | -17 | 0.06 | 8.2 |
| 4628 | 0 | -49 | -11 | 0.15 | 6.9 |
| 6582 | -1 | -127 | -39 | 0.48 | 4.3 |
| 10360 | -1 | -14 | -19 | 0.03 | 7.9 |
| 10476 | +34 | -25 | +3 | 0.14 | 7.6 |
| 10700 | +18 | +29 | +13 | 0.22 | 9.7 |
| 10780 | -13 | -7 | -9 | 0.04 | 7.9 |
| 16160 | -75 | 0 | +34 | 0.20 | 8.6 |
| 20630 | -22 | -4 | -5 | 0.05 | 8.4 |
| 20794 | -67 | -77 | -41 | 0.36 | 5.6 |
| 22049 | -3. | +7 | -20 | 0.09 | 8.8 |
| 26965 | +97 | -13 | -42 | 0.32 | 8.1 |
| 32147 | +1 | -57 | -13 | 0.20 | 6.5 |
| 39587 | +14 | +2 | -7 | 0.10 | 8.6 |
| 100623 | -43 | +22 | +13 | 0.19 | 9.3 |
| 101501 | +8 | -16 | -4 | 0.06 | 7.9 |
| 102365 | -46 | -31 | +9 | 0.21 | 7.1 |
| 109358 | -25 | 0 | +2 | 0.08 | 8.4 |
| 114710 | -47 | +14 | +8 | 0.15 | 8.9 |
| 115617 | -22 | -46 | -32 | 0.15 | 6.9 |
| 128620 | -24 | +10 | +8 | 0.08 | 8.6 |
| 128621 | -20 | +9 | +10 | 0.08 | 8.6 |
| 131156 | +6 | +2 | 0 | 0.07 | 8.5 |
| 149661 | -1 | -1 | -30 | 0.06 | 8.5 |
| 155885 | +0 | -34 | -9 | 0.09 | 7.4 |
| 156274 | +28 | +13 | -24 | 0.19 | 9.3 |
| 165341 | +5 | -19 | -15 | 0.07 | 7.8 |
| 185144 | +36 | +40 | -10 | 0.30 | 10.3 |
| 190248 | -38 | -17 | +3 | 0.11 | 8.1 |
| 191408 | -118 | -53 | +49 | 0.33 | 7.0 |
| 192310 | -69 | -13 | -14 | 0.18 | 8.1 |
| 219134 | -27 | -29 | -3 | 0.18 | 7.1 |

TABLE 3 – The thirteen final candidates for astrobiologically interesting stars: for ease of reference, their Gliese (Gliese & Jahreiss 1979) numbers are also given, besides their *Hipparcos* numbers and names. The stars are ranked by comparing their mass, metallicity, age and Galactic orbit eccentricity to the solar ones. For each parameter similar to the corresponding solar one, we give, qualitatively, in the respective column the symbol "~" and one asterisk is added in the final column, which ranks the stars according to the number of parameters similar to the solar ones. For parameters larger and smaller than solar, the ">" and "<" symbols are given, respectively. For poorly constrained parameters, "?" is given in the corresponding column.

| HD | HIP | Gliese | Name | mass | age | [Fe/H] | ecc. | Rating |
|---|---|---|---|---|---|---|---|---|
| 1581 | 1599 | 17 | ζ Tuc | ~ | ~ | ~ | > | *** |
| 4628 | 3765 | 33 | | < | ? | < | > | |
| 10476 | 7981 | 68 | 107 Psc | < | > | < | > | |
| 16160 | 12114 | 115 | | < | ? | ~ | > | * |
| 32147 | 23311 | 183 | | < | ? | > | > | |
| 100623 | 56452 | 432 | | < | ? | < | > | |
| 102365 | 57443 | 442 | | < | > | < | > | |
| 109358 | 61317 | 475 | β CVn | ~ | ~ | < | ~ | *** |
| 115617 | 64924 | 506 | 61 Vir | ~ | ~ | ~ | > | *** |
| 185144 | 96100 | 764 | σ Dra | < | > | < | > | |
| 190248 | 99240 | 780 | δ Pav | > | ~ | > | ~ | ** |
| 192310 | 99825 | 782 | | < | > | ~ | > | * |
| 219134 | 114622 | 892 | | < | ? | ~ | > | * |

**FIGURE CAPTIONS**

Figure 1 - Observational HR diagram, from *Hipparcos* data, of stars within 10 pc of the Sun. The black box isolates the parameter range +4.0 < $M_V$ < +6.5 and +0.50 < (B-V) < +1.05 (see text) and contains the initial candidates as astrobiologically interesting stars,

Figure 2 – The chromospheric absolute flux in the Ca II HK lines and stellar metallicity of the stars and the Sun. The horizontal dotted lines approximately bracket the range of the active and quiet Sun (Baliunas *et al.* 1995). Stars are seen to divide themselves into active (log R'$_{HK}$ > -4.75) and inactive (log R'$_{HK}$ < -4.75) (Henry *et al.* 1996). Old, magnetically quiet stars are to be found inside the black box; its horizontal limits bracket our cut in metallicity, [Fe/H] < -0.50, and the maximum metallicity found in our selected sample, [Fe/H] ~ +0.4. The star HD 114710 straddles the limit between active and inactive objects, but other data strongly point it to be a young star (see figure 3 and text). Crosses are binary or multiple stars.

Figure 3 – The X-ray luminosities and stellar effective temperatures of the stars and the Sun. Again, stars divide themselves into active, $L_X$ ($10^{27}$) > 10 erg s$^{-1}$, and inactive, below this limit. HD 114710 is seen to have a high X-ray flux: also, its high lithium abundance and fast rotation reveal its youth (see text). The limit for inactive

stars is then drawn at $L_X$ ($10^{27}$) ~ 8 erg s$^{-1}$. Stars are seen to range from F-G (5200 K < $T_{eff}$ < 5900 K) to K-type ($T_{eff}$ < 5100 K) stars: the latter comprise the majority of the sample. Crosses are binary or multiple stars.

Figure 4 – The parameters of the stellar Galactic orbits, the mean Galactocentric distance Rm and the orbital eccentricities. Each of the twelve stars which remained after eliminating the multiple systems, the metal-poor stars, and the near-ZAMS stars are labeled by HD number.

Figure 5 – The U and V velocity components of the full sample. The 1σ and 2σ velocity ellipsoids for thin disk stars with [Fe/H] > -0.4 in the sample of Edvardsson *et al.* (1993), as calculated by Turnbull & Tarter (2003a), are shown; stars are identified as in figure 4.

Figure 6 – The same as figure 5 for the V and W velocity components.

Figure 7 – The star HD 1581 plotted in the theoretical HR diagram. Each evolutionary mass is labeled by its mass, along each track; the full dots alongside the tracks give the ages in Gyr; the luminosity increase $L_{incr}$ undergone by the star since the ZAMS is given, when relevant. The error bars correspond to an uncertainty of 100 K in $T_{eff}$. In luminosity, the average uncertainty is in 0.01 in log $L/L_{\odot}$, and thus the error bars in luminosity are comparable to, or smaller than, the stellar dot size shown.

Figure 8 – Same as figure 7 for HD 4628, 100623 and 102365. Here as well as in the forthcoming diagrams, the evolutionary tracks have been truncated at 10 Gyr, whenever relevant.

Figure 9 – Same as figure 7 for HD 10476 and 185144.

Figure 10 – Same as figure 7 for HD 32147.

Figure 11 – Same as figure 7 for HD 109358.

Figure 12 – Same as figure 7 for HD 115617 and HD 16160.

Figure 13 – Same as figure 7 for HD 190248.

Figure 1 - Porto de Mello et al.
Nearby Astrobiologically Interesting Stars

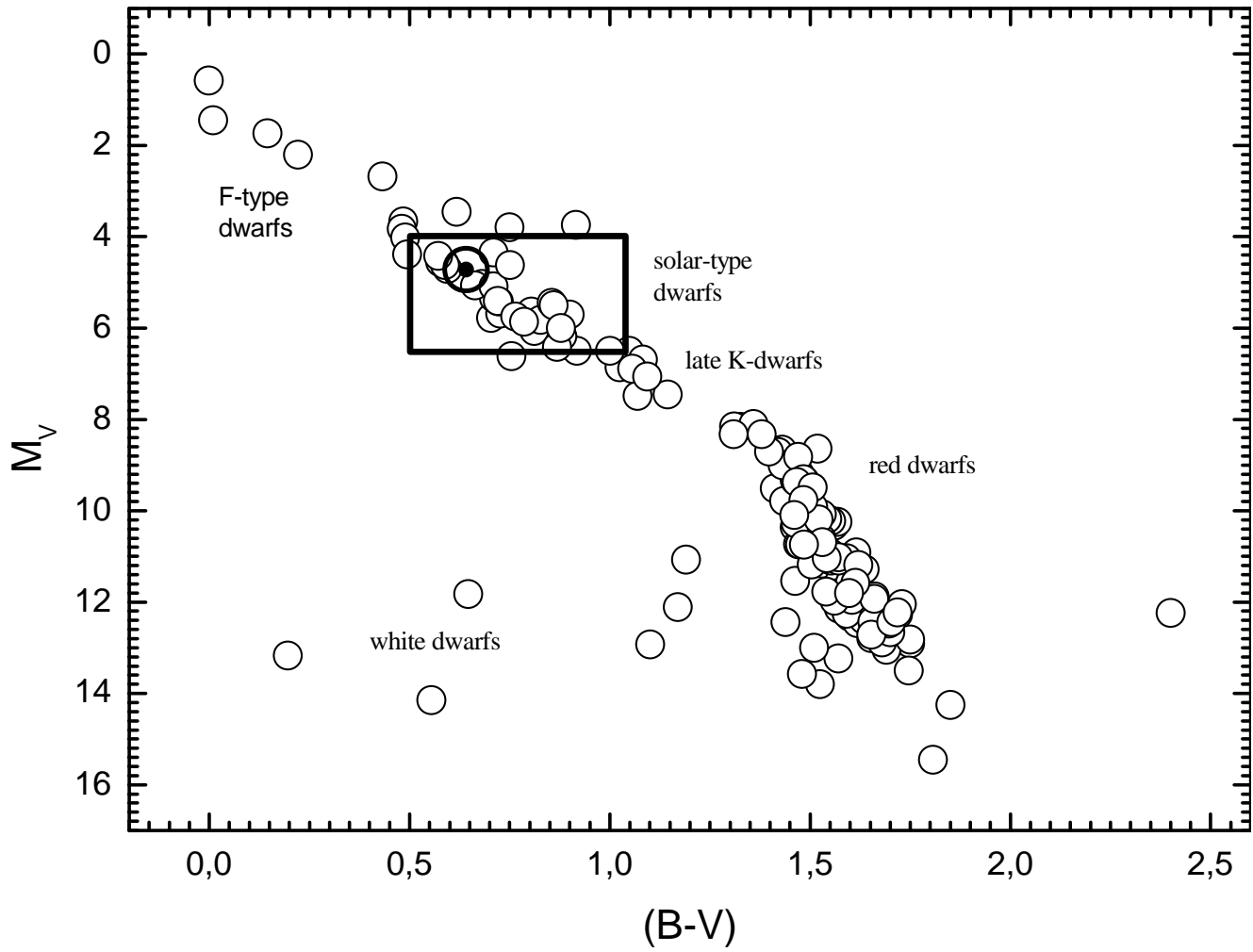

Figure 2 - Porto de Mello et al.
Nearby Astrobiologically Interesting Stars

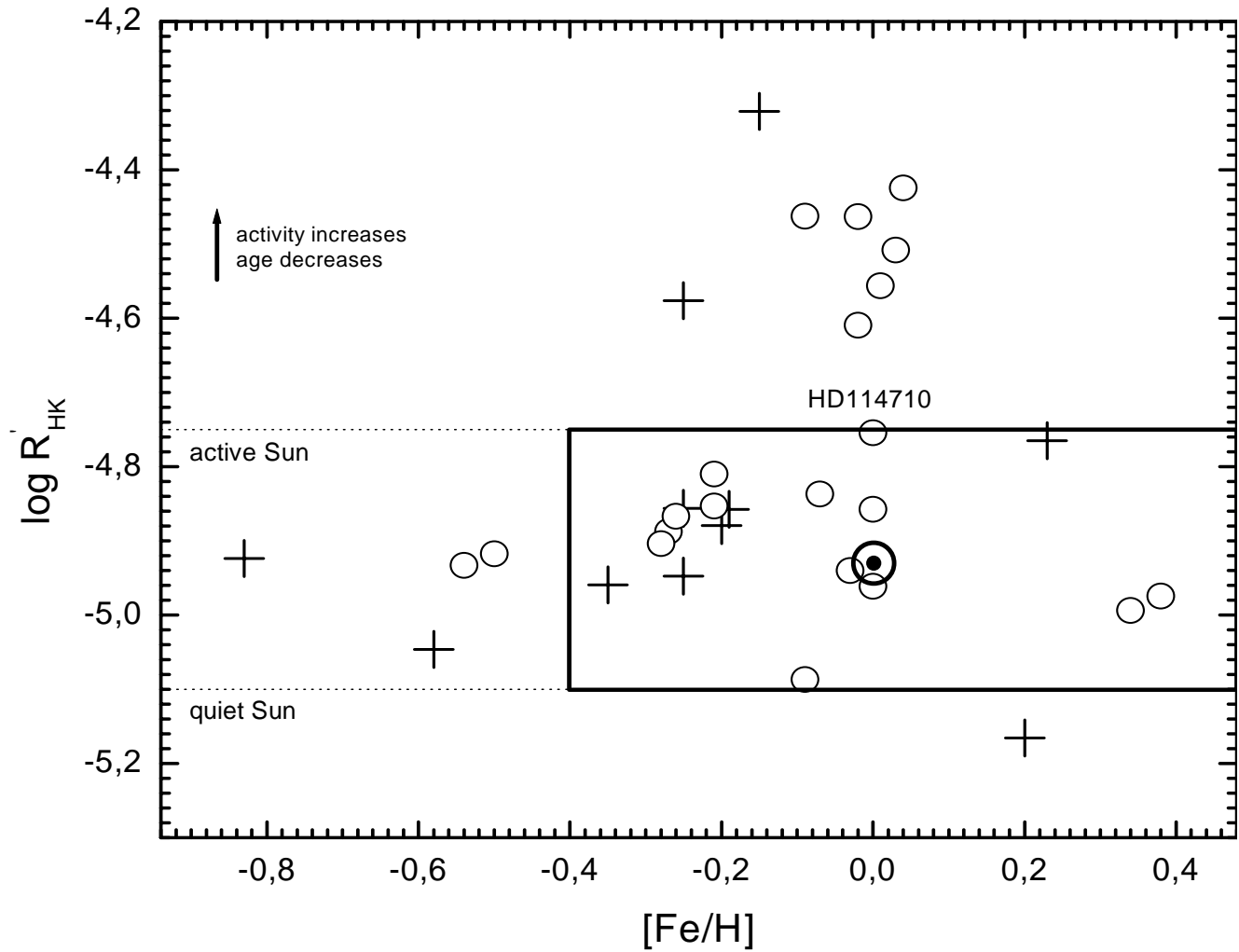

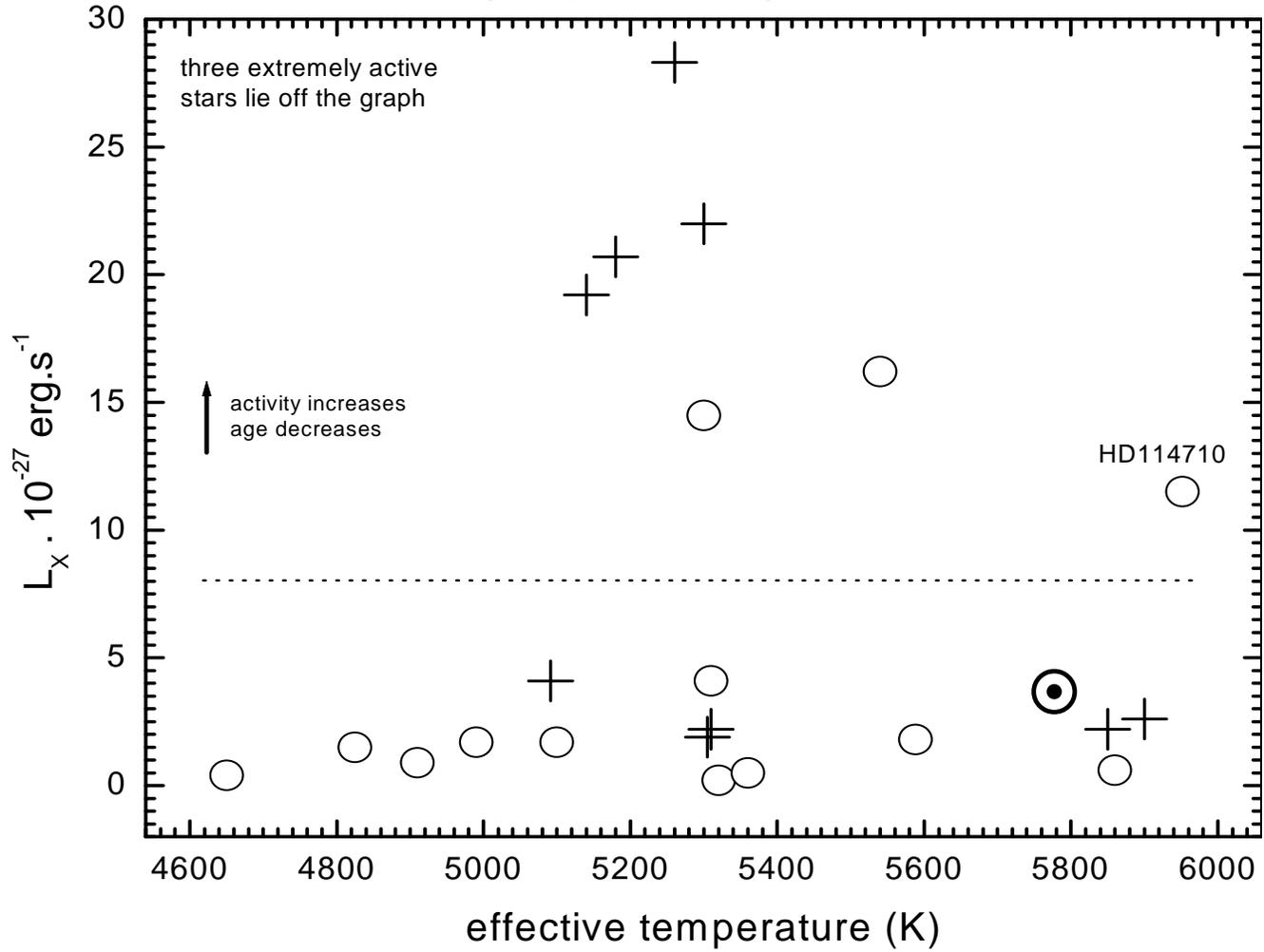

Figure 3 - Porto de Mello et al.
Nearby Astrobiologically Interesting Stars

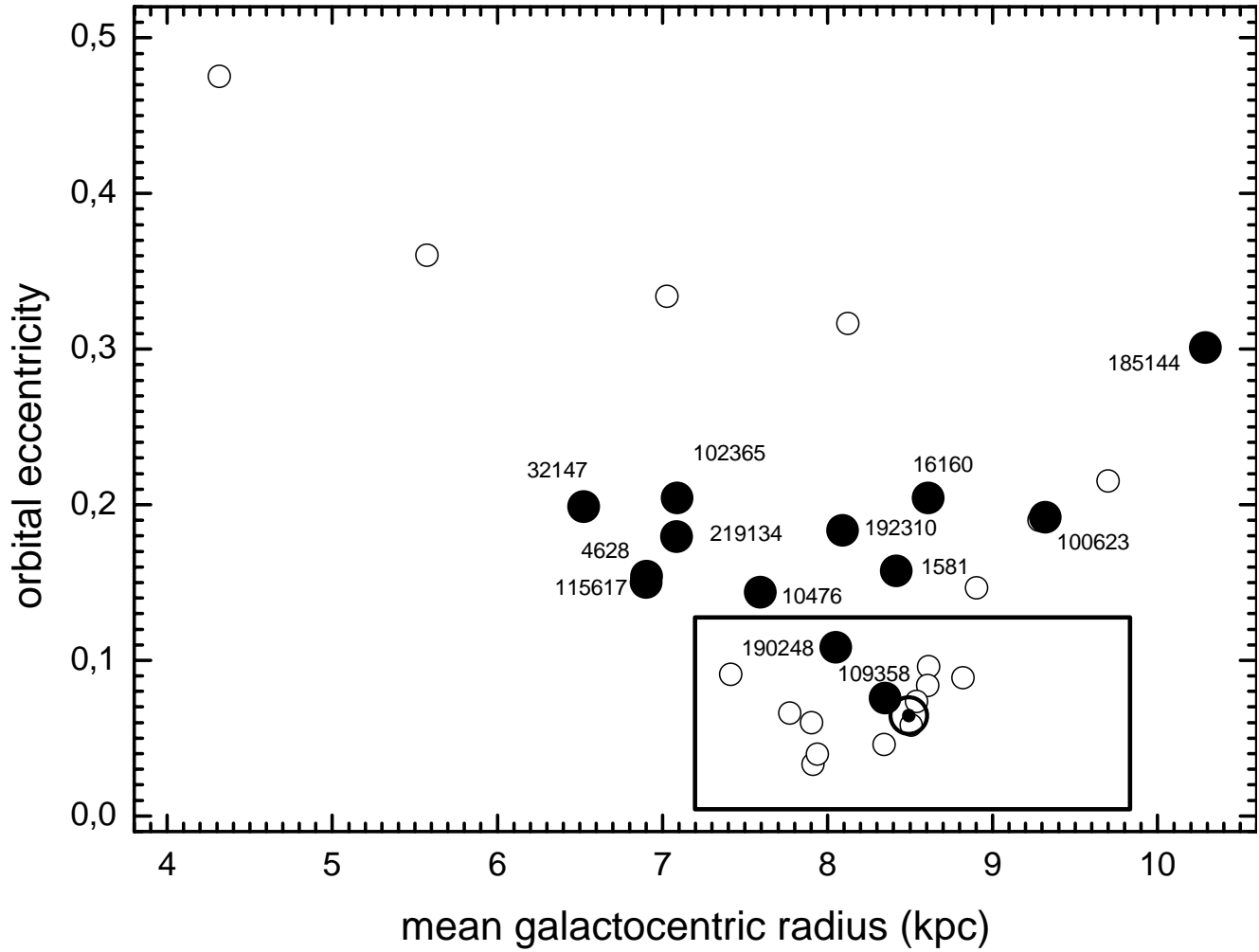

Figure 4 - Porto de Mello et al.
Nearby Astrobiologically Interesting Stars

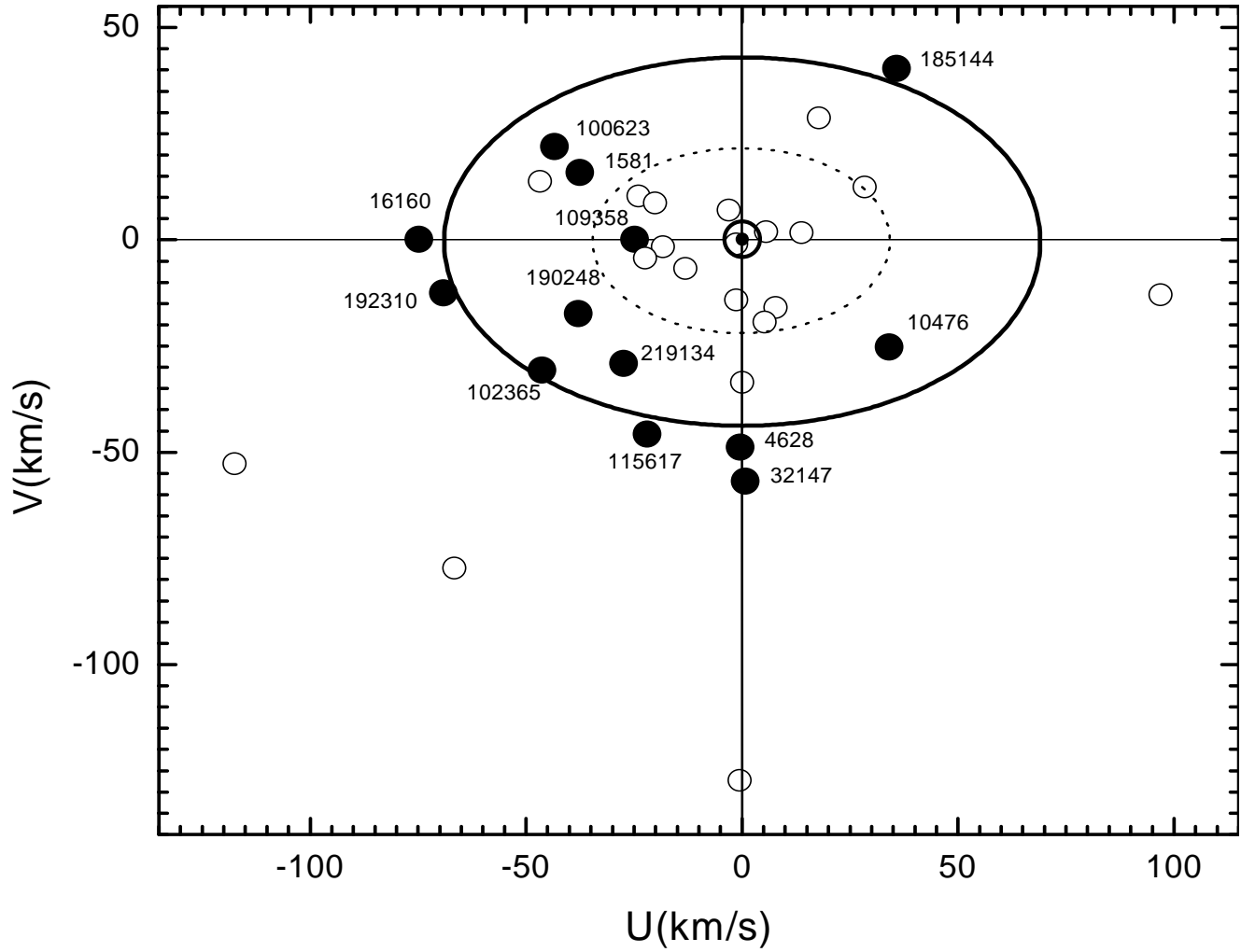

Figure 5 - Porto de Mello et al.
Nearby Astrobiologically Interesting Stars

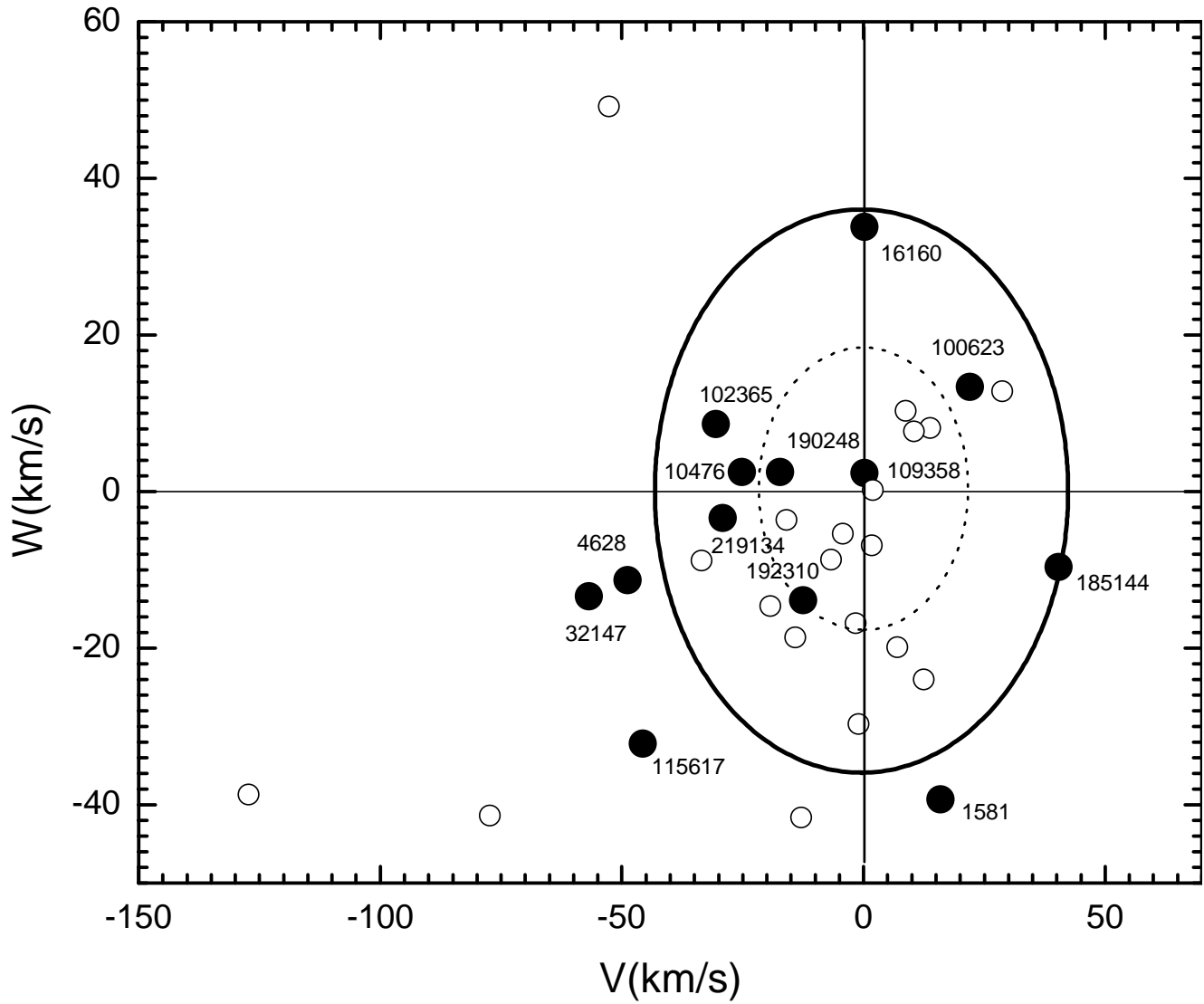

Figure 6 - Porto de Mello et al.
Nearby Astrobiologically Interesting Stars

Figure 7 - Porto de Mello et al.
Nearby Astrobiologically Interesting Stars

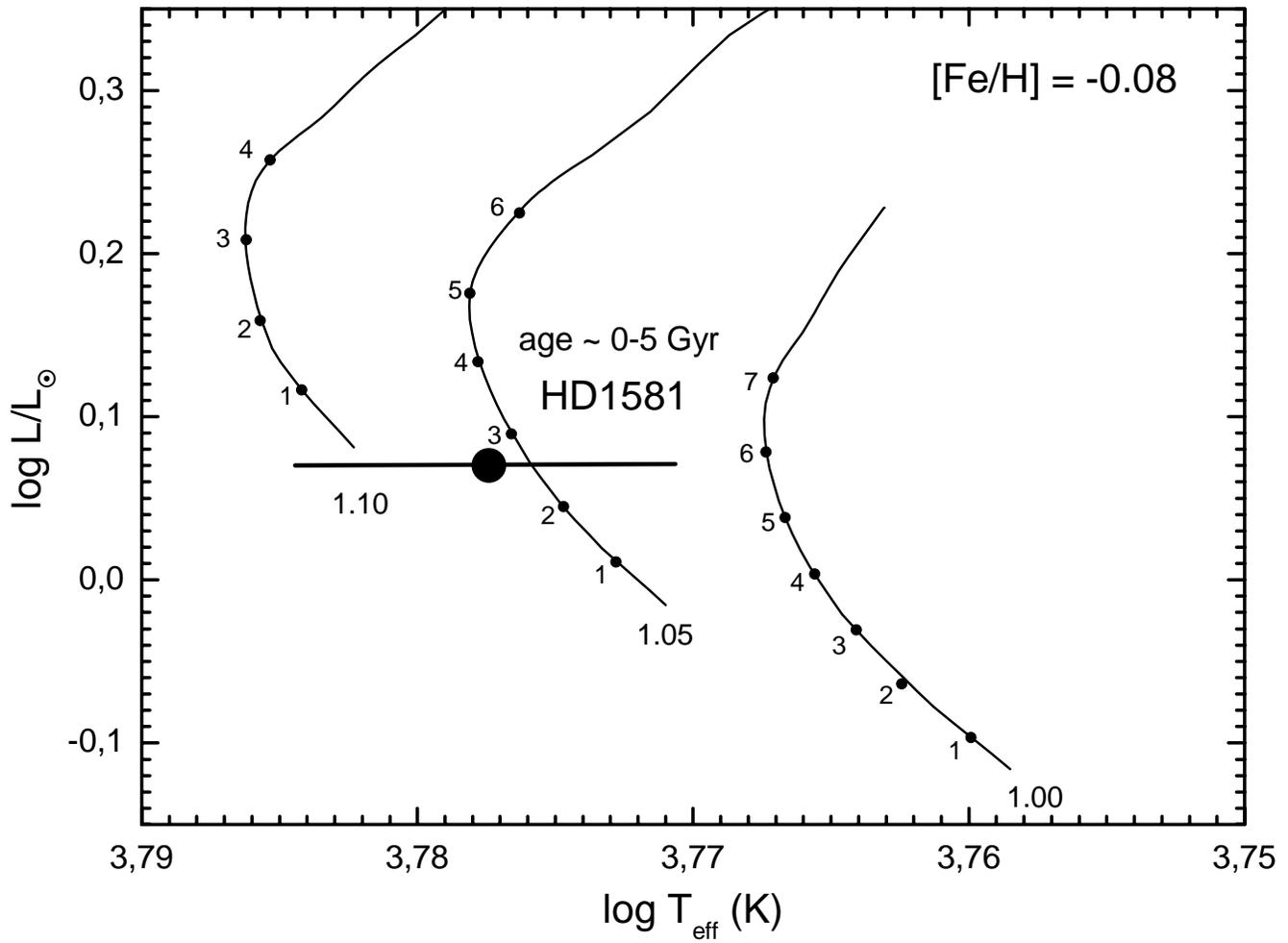

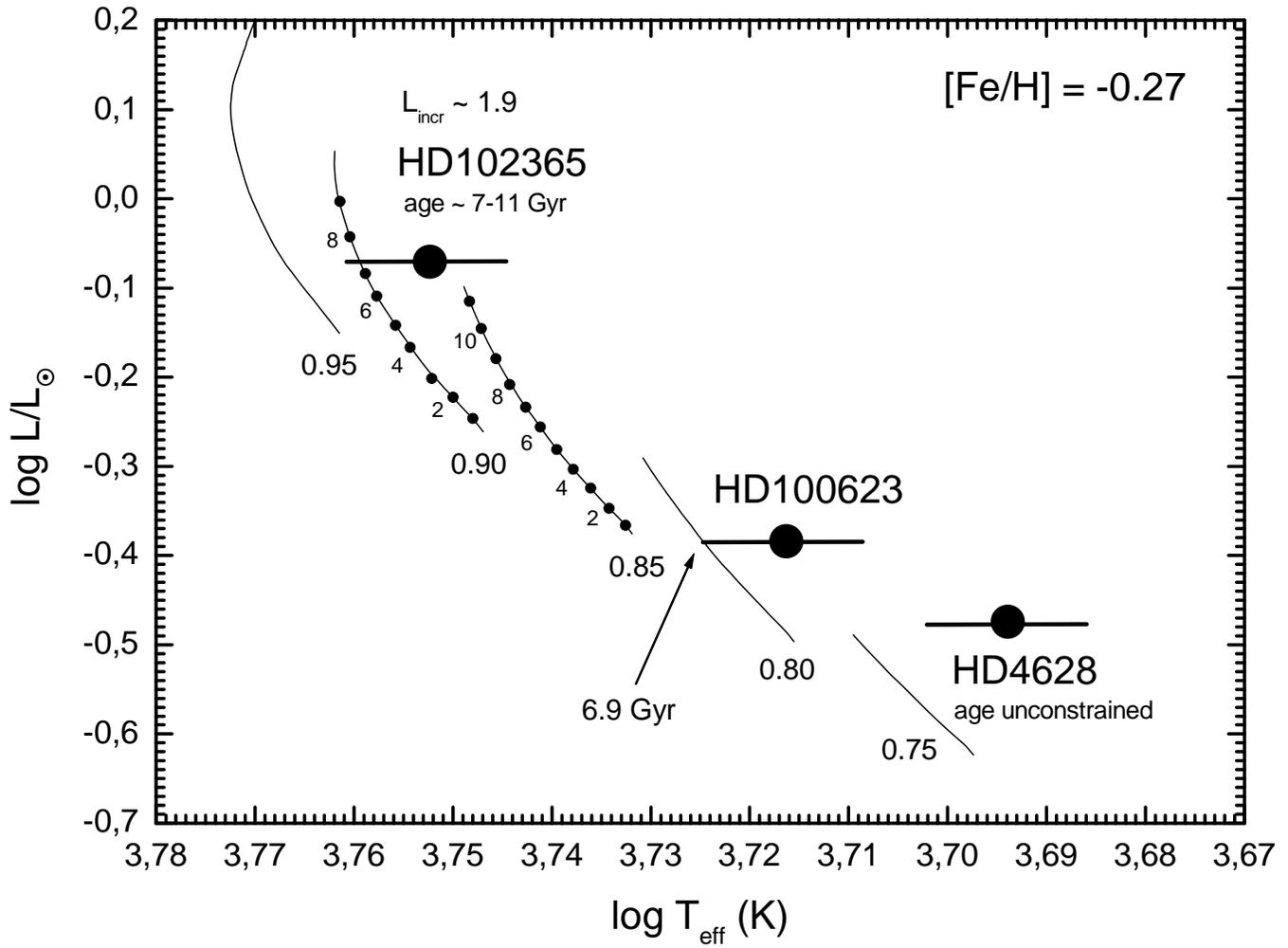

Figure 8 - Porto de Mello et al.
Nearby Astrobiologically Interesting Stars

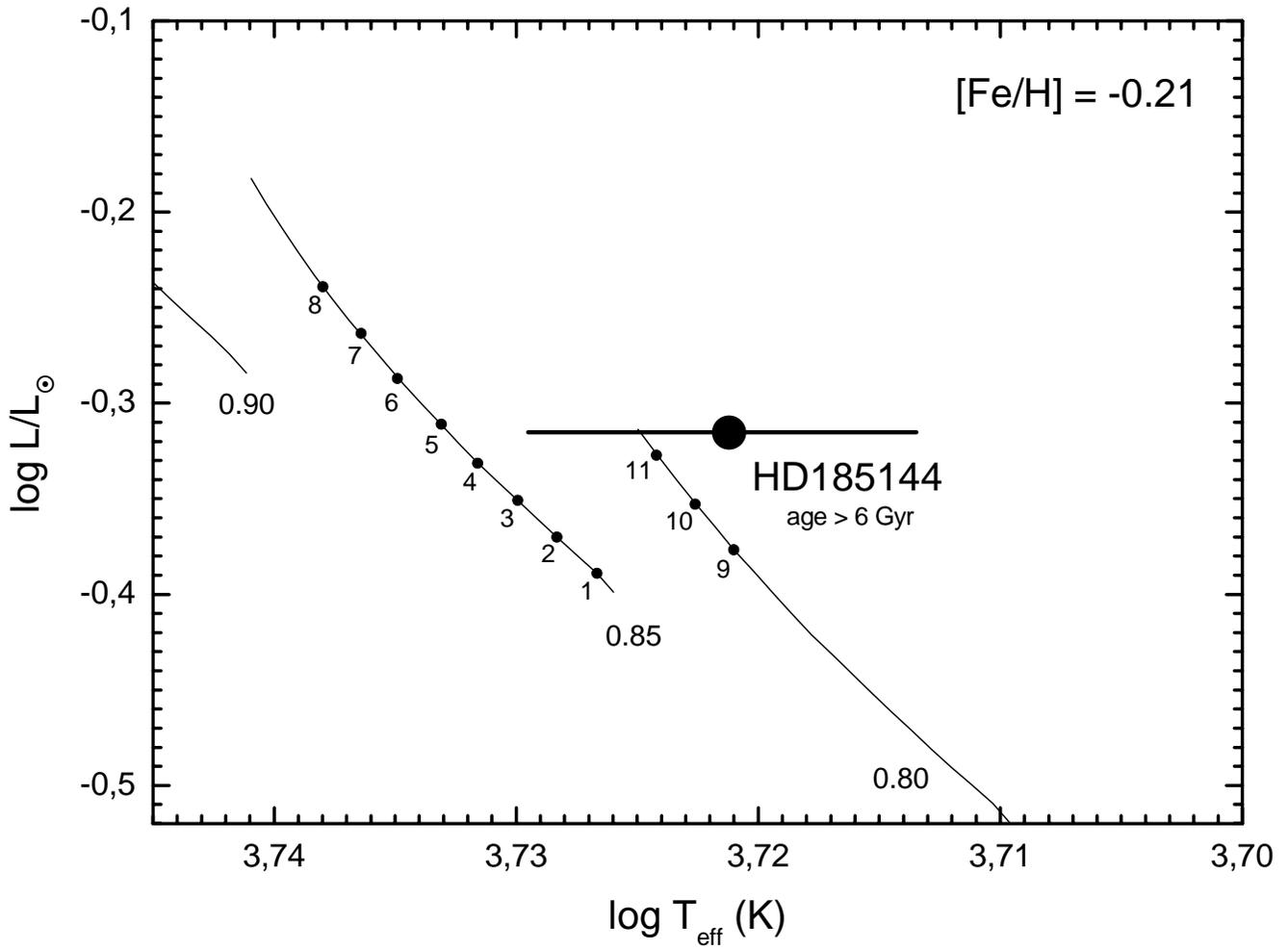

Figure 9 - Porto de Mello et al.
Nearby Astrobiologically Interesting Stars

Figure 10 - Porto de Mello et al.
Nearby Astrobiologically Interesting Stars

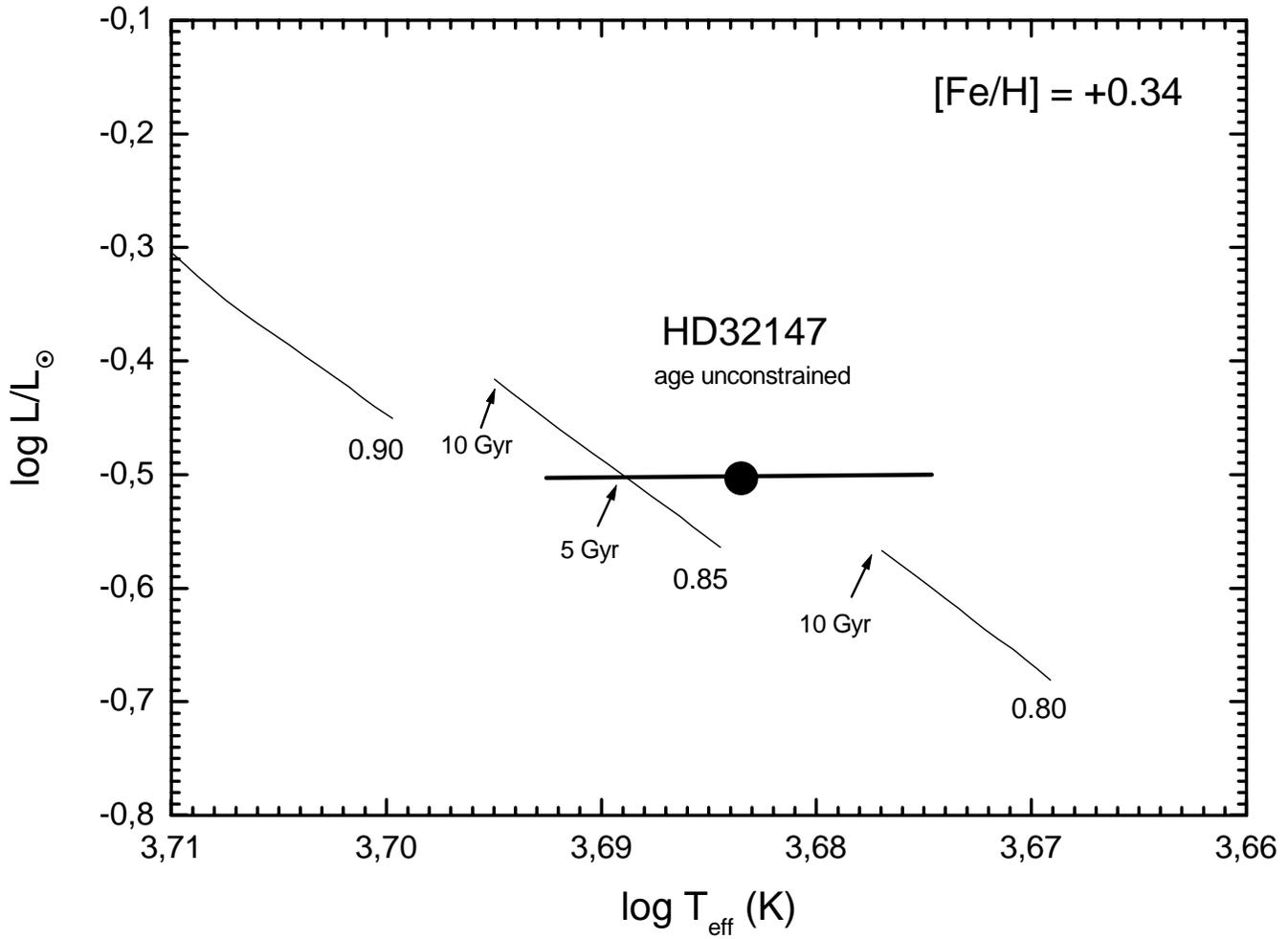

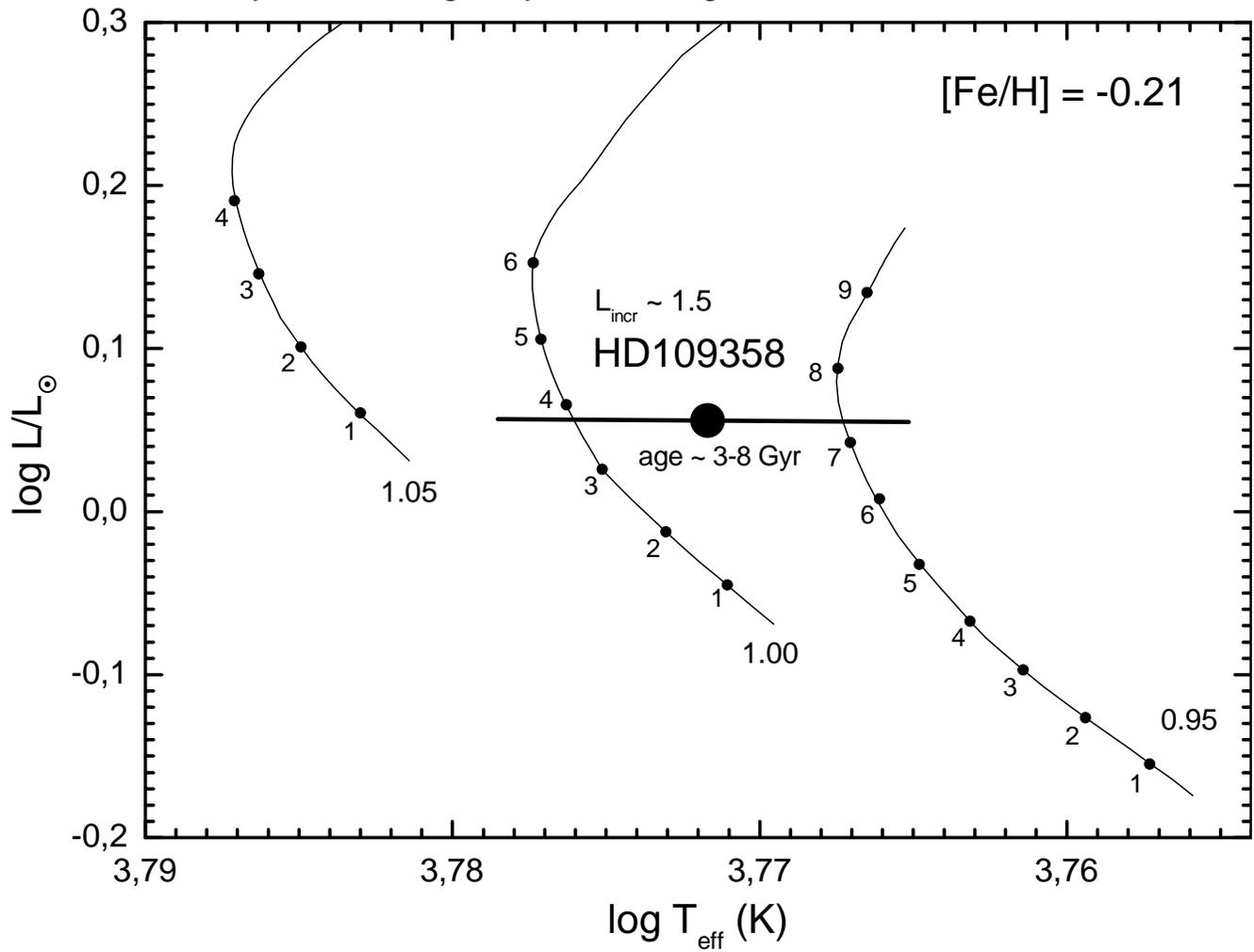

Figure 11 - Porto de Mello et al.
Nearby Astrobiologically Interesting Stars

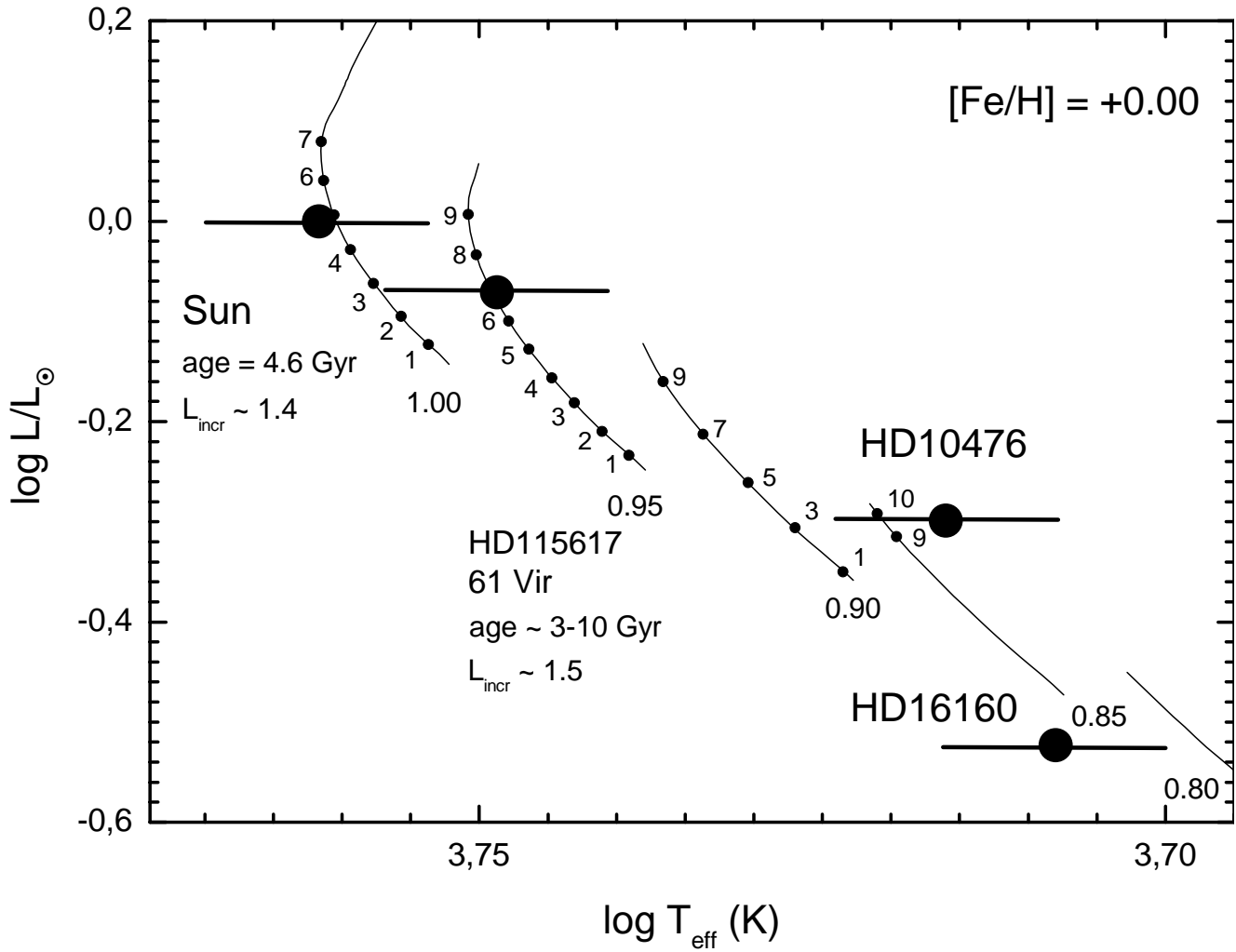

Figure 12 - Porto de Mello et al.
Nearby Astrobiologically Interesting Stars

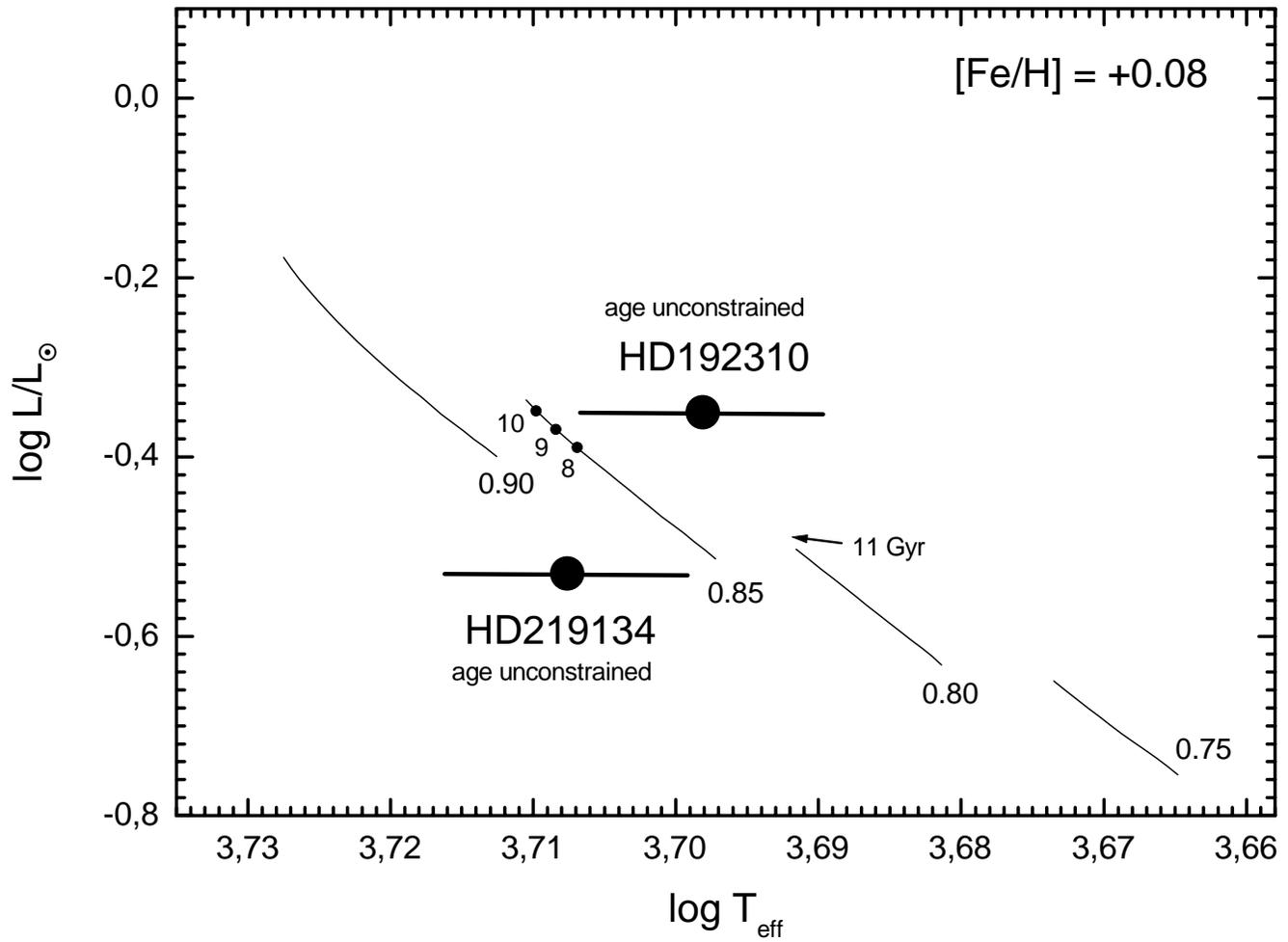

Figure 13 - Porto de Mello et al.
Nearby Astrobiologically Interesting Stars

Figure 14 - Porto de Mello et al.
Nearby Astrobiologically Interesting Stars

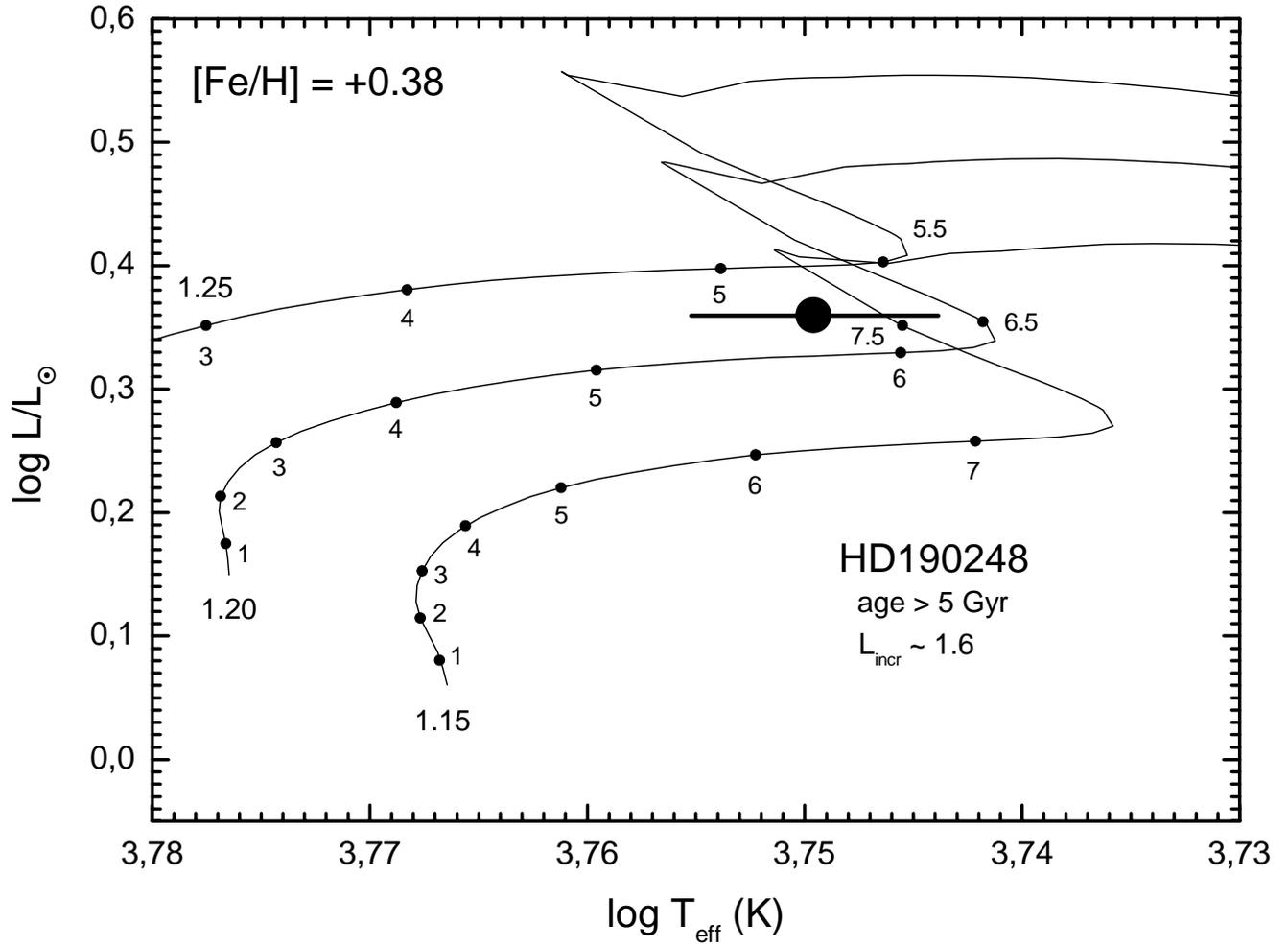

[Fe/H] = +0.38

HD190248
age > 5 Gyr
$L_{incr} \sim 1.6$

log L/L$_\odot$

log T$_{eff}$ (K)